\documentclass[prb,twocolumn,superscriptaddress,nopacs,preprintnumbers,amsmath,amssymb]{revtex4}
\usepackage[utf8x]{inputenc}\SetUnicodeOption{mathletters} \SetUnicodeOption{autogenerated}
\PreloadUnicodePage{4}
\usepackage{graphicx,array,amsmath,ifpdf,amssymb,marvosym,dsfont,xr-hyper,xcolor,url,multirow,cancel,color,mathrsfs,tikz}
\usepackage[OT2,T1]{fontenc}
\usepackage[colorlinks=true,unicode]{hyperref}

\newcommand{\orVomAt}{\vec{\nabla}U_{\mathrm{at}}\cdot [\vec s\times \hat{\vec p}\,]}
\newcommand{\orVomV}{\vec{\nabla}V\cdot [\vec s\times \hat{\vec p}\,]}

\newcolumntype{C}{>{$\displaystyle }c<{$}}
\hypersetup{pdfstartview=FitH,
pdftitle={Spin relaxation in strained silicon},pdflang=en,%
pdfsubject={silicon,multi-valley semiconductor,spin relaxation},%
pdfkeywords={non-perturbative,central cell correction,silicon,multi-valley semiconductor,spin relaxation},%
pdfauthor={Oleg Chalaev,Yang Song,Hanan Dery}}

\begin{document}
\title{Suppressing spin relaxation in silicon}
\author{Oleg~Chalaev}
\affiliation{Department of Electrical and Computer Engineering, University of Rochester, Rochester, New York 14627, USA}
\author{Yang Song}
\affiliation{Department of Electrical and Computer Engineering, University of Rochester, Rochester, New York 14627, USA}
\author{Hanan Dery}
\altaffiliation{hanan.dery@rochester.edu}
\affiliation{Department of Electrical and Computer Engineering, University of Rochester, Rochester, New York 14627, USA}
\affiliation{Department of Physics and Astronomy, University of Rochester, Rochester, New York 14627, USA}
\begin{abstract}
Uniaxial compressive strain along the [001] direction strongly suppresses the spin relaxation in silicon. When the strain level is large enough so that electrons are redistributed only in the two valleys along the strain axis,  the dominant scattering mechanisms are quenched and electrons mainly experience intra-axis scattering processes (intravalley or intervalley scattering within valleys on the same crystal axis). We first derive the spin-flip matrix elements due to intra-axis electron scattering off impurities, and then provide a comprehensive model of the spin relaxation time due to all possible interactions of conduction-band electrons with impurities and phonons. We predict nearly three orders of magnitude improvement in the spin relaxation time of $\sim10^{19}\text{cm}^{-3}$ antimony-doped silicon (Si:Sb) at low temperatures.  
\end{abstract}
\maketitle 

\section{Introduction}



Silicon is a promising material choice for spintronic devices that require long spin lifetimes.\cite{Zutic_RMP04,Zutic_PRL06,Dery_PRB06,Fabian_APS07,Dery_Nature07,Behin_NatureNano10,Song_PRB10,Sverdlov_PhysRep15,Hamaya_bookChapter,Wen_PRAp16}  When electrons reach their saturation drift velocity, spin information in silicon can be transported over hundreds of microns,\cite{Huang_PRL07,Huang_APL08,Li_PRL12,Qing_PRB15} being compatible with on-chip interconnect length scales.\cite{Zutic_NatMater11,Dery_APL11} Transport over such distances is possible not only due to the relatively weak spin-orbit coupling of silicon atoms, but also owing to two major manifestations of the crystal space inversion symmetry. The first one is the spin degeneracy of the energy bands in centrosymmetric materials resulting in cancelation of the Dyakonov-Perel spin relaxation mechanism.\cite{Dyakonov_JETP71} The second manifestation, first studied by Yafet, is that space inversion and time reversal symmetries weaken the electron's spin-flip scattering amplitude due to interaction with phonons when $|\mathbf{k}_i \pm \mathbf{k}_f|a \ll 1$.\cite{Yafet_SSP63,Cheng_PRL10,Li_PRL11,Song_PRB12} Here, $\mathbf{k}_{i(f)}$ is the initial (final) electron's wavevector and $a$ is the lattice constant. In one case the electron remains in the same valley, $|\mathbf{k}_i - \mathbf{k}_f|a \ll 1$, and in the other it is scattered to the opposite valley, $|\mathbf{k}_i + \mathbf{k}_f|a \ll 1$. The former denotes intravalley scattering and the latter intervalley scattering between opposite valleys, termed $g$-process in silicon.\cite{Long_PR60,Streitwolf_PSS70,Ridley_Book,Yu_Cardona_Book} These weak spin-flip scattering processes can be generally classified as `intra-axis' scattering if the conduction-band edge is degenerate having more than one lowest-energy valley in the Brillouin zone. 

Contrary to the weak intra-axis spin flip process in centrosymmetric materials, `inter-axis' valley scattering can have a much larger spin-flip amplitude.\cite{Song_PRL14,Li_PRL11} In this type of intervalley scattering, termed $f$-process in silicon,\cite{Long_PR60,Ridley_Book,Yu_Cardona_Book} the initial and final valleys  are not connected by time reversal or space inversion symmetries. The electron transition between the valleys is mediated by interaction with shortwave phonons or short-range scattering off impurities.\cite{Li_PRL11,Song_PRL14} While the contribution from shortwave phonons becomes negligible at low temperatures due to their large energy compared with the thermal energy ($k_BT$), spin relaxation due to impurities is unavoidable at all temperatures. In either case, inter-axis valley scattering is the dominant means to relax the spins of itinerant electrons in unstrained silicon or germanium.\cite{Li_PRL11,Song_PRB12,Li_PRL12,Song_PRL14,Li_PRB12,Tang_PRB12,Guite_APL12,Li_PRL13,Lohrenz_PRB14,Giorgioni_APL14,Yu_JPCM15,Dushenko_PRL15}

Suppression of the detrimental inter-axis valley scattering can be achieved by applying uniaxial compressive strain along the [001]-crystallographic direction in silicon.\cite{Dery_APL11,Tang_PRB12,Ghosh_Micro15,Osintsev_SSE15,Yang_arXiv16} This strain configuration lowers the energy edge for the pair of valleys along the strain axis, while the energy edge of all other valleys is raised. For the valley splitting energy to be sufficiently large compared with $k_BT$, the
strain levels should be of the order of 0.1\% at 30~K and 1\% at room temperature. Under these conditions, electrons populate the low-energy valley pair, and therefore can only experience intra-axis scattering processes (intravalley and $g$-process). To date, there are no theoretical models that quantify the spin relaxation due to intra-axis electron scattering off impurities.  The aim of this paper is to fill this missing component,  and to compare the relative contributions of electron-phonon and electron-impurity interactions to spin relaxation.

Additional motivation to our work stems from the need to improve the electrical spin injection from ferromagnetic metals to heavily doped $n$-type silicon.\cite{Min_NatureMaterials06,Appelbaum_Nature07,Jonker_NatPhys07,Erve_APL07,Jang_PRB08,Mavropoulos_PRB08,Dash_Nature09,Erve_IEEE09,Jang_PRL09,Ando_APL09,Sasaki_APE09,Li_APL09b,Grenet_APL09,Kioseoglou_APL09,Huang_PRB10,Sasaki_APL10,Sasaki_IEEE10,Ando_APE10,Jansen_PRB10,Sasaki_APE11,Lu_PRL11,Ando_APL11b,Breton_Nature11,Jeon_APL11,Li_NatComm11,Sasaki_APL11,Ando_APL11a,Jansen_NatMater12,Shikoh_PRL13,Pu_PRL15,Lee_APL16} Because of the so-called conductivity and spin lifetime mismatch problem,\cite{Schmidt_PRB00,Rashba_PRB00,Fert_PRB01} electrical spin injection to semiconductors is largely limited to tunneling or ballistic injection of hot electrons.\cite{Hanbicki_APL03,Crooker_Science05,Shiraishi_PRB11,Ando_APL11,Chang_SST13,Erve_NatComm15,Liu_NatComm16,Txoperana_JPDAP16,Monsma_PRL95,Jansen_JPDAP03,Huang_JAP07,Li_APL08,Li_APL09a,Lu_APL13} In the former, electrons tunnel across the built-in potential barrier in the semiconductor side of the junction. The barrier is formed by the depletion region, existing both in direct metal-semiconductor Schottky contacts and metal-oxide-semiconductor structures with ultrathin oxide layers. Effective tunneling requires interface doping with donor concentrations of 10$^{19}$~cm$^{-3}$ or higher so that the thickness of the depletion region is only a few nanometers.\cite{Hanbicki_APL03,Crooker_Science05,Shiraishi_PRB11,Ando_APL11} Such degenerate doping levels come with a penalty of enhanced spin relaxation due to electron-impurity scattering.\cite{Lepine_PRB70,Gershenzon_PSS70,Ue_PRB71,Quirt_PRB72,Gershenzon_PSS72a,Gershenzon_PSS72b,Onda_JPSJ73,Chazalviel_JPCS74,Pifer_PRB75,Ochiai_PSS76,Zarifis_PRB87,Zarifis_PRB98} As we will show, application of strain to suppress the dominant inter-axis valley scattering greatly increases the spin lifetime due to a much weaker effect of intra-axis scattering off impurities.

This paper is organized as follows. Section~\ref{sec:back} provides a theoretical background starting with a summary of our previous findings on inter-axis valley scattering in multivalley semiconductors. The second part of Sec.~\ref{sec:back} contains a discussion on a compact $\mathbf{k}$$\cdot$$\mathbf{p}$ Hamiltonian whose eigenstates are used in  Sec.~\ref{sec:derivation} to derive the intra-axis spin-flip matrix elements due to scattering off impurities. To keep the discussion succinct, many  technical details on the derivation of the Hamiltonian and its utilization in the context  of intra-axis spin flips are provided in Appendix~\ref{sec:symmetries}. Section \ref{sec:relax} includes integration of all spin-flip processes due to electron-phonon and electron-impurity interactions, from which one can quantify the dependence of the spin relaxation on strain, temperature, donor impurity concentration, and donor identity. Section \ref{sec:result} includes results and discussion of these dependencies, and conclusions are provided in Sec. \ref{sec:conc}. Finally, Appendix ~\ref{App:Fterms} includes technical computational details.

\section{Theoretical Background} \label{sec:back}

Contrary to single-valley semiconductors, such as GaAs, a zero-velocity wave packet can scatter off an impurity in multivalley semiconductors. The reason is that in addition to velocity, the wave packet has an extra ``quantum number'' -- the valley index. We have recently shown that the spin relaxation due to scattering off impurities is governed by the inter-axis valley change, while the velocity of the wave packet plays a minor role.\cite{Song_PRL14}  This spin-flip process is enabled by short-range interaction of the scattered electron with the spin-orbit coupling potential of the impurity. To quantify the scattering amplitude, one can make use of the analytical relation between the phase shift of scattered states and the binding energies of the impurity bound states.\cite{Baldereschi_PRB70,Gantmakher_Levinson_Book,Ralph_PRB75} This relation bypasses the need to rely on ab-initio calculations in order to quantify the spin relaxation. Consequently, we were able to quantify the spin-flip scattering amplitude by using empirical fine-structure parameters of the impurity bound states. The spin-flip matrix element due to inter-axis valley scattering off an impurity can then be written as,\cite{Song_PRL14}
\begin{equation}  
M_{sf} = C \pi \delta_{B} \Delta_{so}\,,\qquad \delta_B = \frac{a_B^3}{V}, \label{eq:M_appr}
\end{equation}
where $C$ is a constant of the order of unity that depends on spin orientation.  $V$ and $a_{B}$ are the crystal volume and the electron Bohr radius, respectively.  $\Delta_{so}$ is the spin-splitting of the ground-state impurity level whose amplitude is commensurate with the central cell correction coming from the difference between the spin-orbit interaction of the impurity and silicon atoms, $\delta V_{so}$. Because the Bohr radius is large compared with the impurity's core-region radius wherein $\delta V_{so}$ is relatively strong ($>1$~nm vs $\sim$$\,$0.1~nm), the central cell correction has to be strong enough in order to produce measurable spin-splitting. In other words, the strength of $\delta V_{so}$ has to compensate its short range. The dependence of $\Delta_{so}$ on the difference between the spin-orbit interaction of the impurity and silicon atoms is indeed corroborated in experiments, showing that $\Delta_{so}$ is of the order of 0.3, 0.1, and 0.03~meV for Sb, As, and P impurities, respectively.\cite{Aggarwal_PR65,Castner_PR67} 

To better understand the details of the impurity's spin splitting, we consider a substitutional impurity atom surrounded by four silicon atoms in a tetrahedral molecular geometry. The vast majority of shallow donors in Si are represented by such substitutional impurities whose potential has $T_d$ point-group symmetry. Due to the valley-orbit coupling within the $T_d$ impurity, the ground state level (1$s$) is split into spin-independent nondegenerate ($A_1$), doubly degenerate ($E$)  and triply degenerate ($T_2$) states where the overall 6-fold multiplicity comes from the number of conduction edge states (valley centers).\cite{Kohn_SSP57,Ramdas_RPP81}  $A_1, E$ and $T_2$ denote the symmetrized linear combinations of these valley edge states  under $T_d$ group operations. When adding the spin degree of freedom, the notable measured effect from the spin-orbit coupling is attributed to the spin splitting of the triply degenerate state ($T_2$).\cite{Castner_PR67,Song_PRL14} This splitting is denoted by $\Delta_{so}$ in  (\ref{eq:M_appr}). We note that $\Delta_{so}=0$ for direct band-gap semiconductors, in which substitutional donors do not change the point-group symmetry and their ground state is non-degenerate (the conduction band has one low-energy valley in the zone center). 

\subsection{$X$-point $\mathbf{k}$$\cdot$$\mathbf{p}$ Hamiltonian} 

When the inter-axis scattering is quenched by strain, various types of weak intra-axis mechanisms become relevant. Contrary to the inter-axis mechanisms, the intra-axis ones depend on the velocity of the wave packet. The spin-flip dependence on the momentum of the incoming and scattered wave packets can be captured by employing a spin-dependent $\mathbf{k}$$\cdot$$\mathbf{p}$ Hamiltonian to describe the low-energy conduction-band states. We construct the Hamiltonian by using its invariance to the symmetry operations of the space group $G^2_{32}$, which describes the symmetry of the $X$ point at the edge of the Brillouin zone in diamond crystal structures.\cite{Yu_Cardona_Book,Hensel_PR65,Jones1960,Dresselhaus_PR67,Dresselhauses_Jorio_Book,Bradley_Cracknell2010} In silicon, the $X$ point is closer to the absolute conduction band minimum than all other high symmetry points, thereby allowing us to reliably express the low-energy  conduction states using a minimal set of basis functions.\cite{Li_PRL11,Hensel_PR65} Due to the symmetry of the crystal, only one of the six conduction band valleys in silicon is studied, and we arbitrary identify it as the valley along the $+z$ crystallographic axis for which the $X$ point corresponds to $\mathbf{k} = (0,0,2\pi/a)$. The results presented below can be readily extended to other valleys by cyclic coordinate permutations.  

The $\mathbf{k}$$\cdot$$\mathbf{p}$ state expansion in the vicinity of the $X$-point is carried by employing basis functions for the lowest pair of conduction bands and upper pair of valence bands. Inclusion of the valence states is imperative since they bring in the mass anisotropy and spin mixing of the states.\cite{Hensel_PR65,Li_PRL11,Song_PRB12} The nomenclature for the irreducible representations (IRs) of the conduction and valence pairs is $X_1$ and $X_4$, respectively. Each of these IRs is two-dimensional due to the twofold band degeneracy at the $X$ point of diamond crystal structures, originating from time-reversal and glide-reflection symmetries.\cite{Yu_Cardona_Book}  We  denote the corresponding spinless basis states as $X_1=(X_1^{2'},X_1^1)$ and  $X_4=(X_4^x,X_4^y)$. The superscript indexing of the basis states is reminiscent of their compatibility relations with IRs of the  $\Delta$-axis connecting the $\Gamma$ and $X$ points. Namely, $\Delta_{2'(1)}$ denotes the top (bottom) branches of the conduction band, while $\Delta_{x,y}$ denote the degenerate valence band along the $\Delta$ axis (heavy and light holes). The compatibility relation also allows us to relate the basis functions along the $\Delta$-axis. For example, $\psi_{\Delta_{2'},k_z}(\mathbf{r}) \simeq e^{i(k_z-k_X)z} \psi_{X_1^{2'}}(\mathbf{r})$ where $k_X = 2\pi/a$. Following the notation of Ref.~[\onlinecite{Hensel_PR65}], the basis components are chosen to be complex conjugate of each other
\begin{equation}  \label{eq:8}
X_1^{2'*}=X_1^1,\quad X_4^{x*}=X_4^y.
\end{equation}

While our expansion is carried around one of the equivalent $X$ points, the eigenstates involved in an intervalley $g$-process can be readily connected by time-reversal and space inversion symmetries, \begin{eqnarray}  \label{eq:5}  
 |\uparrow \mathbf{k}\rangle &=& -\hat I|\uparrow -\mathbf{k}\rangle ,\quad \,\,\,    |\downarrow \mathbf{k}\rangle =-\hat I|\downarrow -\mathbf{k}\rangle\,, \nonumber \\
 |\downarrow \mathbf{k}\rangle &=& -i\hat T|\uparrow -\mathbf{k}\rangle ,\quad  |\uparrow \mathbf{k}\rangle =i\hat T|\downarrow -\mathbf{k}\rangle\, ,
\end{eqnarray}
$\hat I$ and $\hat T$ are operators of space and time reversal. These relations are derived by choosing a gauge according to which spatial inversion negates spinors, and by letting complex conjugation and spatial inversion to act on basis functions~\eqref{eq:8} equivalently.

Adding the spin-orbit coupling at the  $X$-point, we note that the IRs of the corresponding double group cannot be factorized into product of $\{X_1^{2'},X_1^1,X_4^x,X_4^y\}$ and $\{\uparrow ,\downarrow\}$.\cite{Li_PRL11} As a consequence, even in the absence of impurities and at $\mathbf{k}=(0,0,k_X)$, the $\mathbf{k}$$\cdot$$\mathbf{p}$ Hamiltonian contains non-zero interband spin-mixing terms (see Appendix~\ref{sec:symmetries} or Ref.~[\onlinecite{Song_PRB12}]),
\begin{equation}  \label{eq:21}
      H_{cv}\Big|_{k=0}=i\Delta _X(\rho _x\otimes\sigma _y-\rho _0 \otimes\sigma _x).
\end{equation}
$\rho _x,\rho _y,\rho _z$  are pseudospin Pauli matrices due to the twofold band degeneracy in the $X$-point,  and $\sigma _x,\sigma _y,\sigma _z$ are the spin
Pauli matrices;  $\rho_0$ is a $2\times 2$  orbital unity matrix.
$\Delta_X \simeq$~ 4~meV is the finite spin-orbit coupling parameter between $X_1$ and $X_4$.\cite{Li_PRL11,Song_PRB12} The presence of the spin-mixing term~\eqref{eq:21} in the Hamiltonian means that fully polarized waves are not eigenstates of an impurity-free Hamiltonian. Its eigenstates are slightly spin-mixed,
\begin{eqnarray}  \label{eq:48} 
    \langle \mathbf{r} | \Uparrow      \mathbf{k} \rangle &=& e^{i\mathbf{k}\mathbf{r}}\sum ^4_{i=1}[\uparrow A_i(\mathbf{k})+\downarrow B_i(\mathbf{k})]Y_i(\mathbf{r}), \nonumber \\
    \langle \mathbf{r} | \Downarrow \mathbf{k} \rangle &=& i\hat I\hat T \langle \mathbf{r}\, |\Uparrow \mathbf{k} \rangle, 
\end{eqnarray} 
where $(Y_1,Y_2,Y_3,Y_4)=(X_1^{2'},X_1^1,X_4^x,X_4^y)$, and
\begin{equation}\label{eq:a1}  \begin{split}
\mathbf{A}(\mathbf{k}) = &\left[0,\,1,\,-\frac{k_xP}{E_g},\,-\frac{k_yP}{E_g}\right],\\
\mathbf{B}(\mathbf{k}) = &\left[\frac{\eta (k_x-ik_y)}{\Delta _c},\,0,\,-\frac{\Delta '_X}{E_g},\,-\frac{i\Delta '_X}{E_g}\right],\\
\eta =\frac{2iP\Delta _X}{E_g},&\ \Delta '_X=\Delta _X-\alpha k'_z,\ \displaystyle \Delta _c=\frac{2\hbar ^2k_0k_z'}m.
\end{split}\end{equation}
$k'_z\equiv k_z-k_X$ and $\Delta _c$ are negative for the $+z$~valley. The notations of constants in~\eqref{eq:a1} and their values are the same as in Ref.~[\onlinecite{Song_PRB12}]: $E_g \simeq$~4.3~eV is the $X$-point band gap, $P\simeq$~10~eV$\cdot$$a/2\pi$ is the interband momentum matrix element where $a$=5.43~\AA~in Si, and $|\Delta_c| \simeq$~0.5~eV is the energy splitting between the top and bottom conduction bands at the valley-edge position (15\% away from the $X$ point toward the $\Gamma$ point; $k_0 = 0.15k_X$).  Finally, $\alpha  \simeq -3.1$~meV$\cdot$$a/2\pi$  is a correction to the $X$-point spin-orbit coupling parameter ($\Delta_X$) due to the finite distance of the valley bottom from the $X$ point. Below, we make use of the state expansion in (\ref{eq:a1}) to derive intra-axis spin-flip matrix elements. 

\section{ Spin-flip processes in strained silicon due to scattering off impurities} \label{sec:derivation}

A central goal of this paper is to derive the intra-axis matrix elements due to scattering off impurities. These matrix elements, $M_{sf}(\mathbf{k}_i,\mathbf{k}_f) = \langle \Downarrow \mathbf{k}_f | V |\Uparrow \mathbf{k}_i\rangle$ where $V$ is the impurity potential, govern the spin relaxation when the strain-induced valley splitting is large enough to quench inter-axis scattering. For elastic scattering off impurities, $k_i=k_f$, the resulting spin relaxation rate is\cite{Song_PRB12}
\begin{equation} \label{eq:tau_s}
      \frac{1}{\tau_s} = \frac{4\pi N_DV}{\hbar}  \Big\langle \sum_{\mathbf{k}_f}  |M_{sf}(\mathbf{k}_i,\mathbf{k}_f)|^2 \delta(E_{\mathbf{k}_i}-E_{\mathbf{k}_f}) \Big\rangle_{\mathbf{k}_i}\!,\,\,\,\,
\end{equation}
where $N_D$ is the donor impurity concentration. The average over $\mathbf{k}_i$ represents weighted integration over $\partial \mathcal{F} / \partial E_{\mathbf{k}_i}$ where $\mathcal{F}$ denotes the electron distribution function. This weighted integration is exact for any distribution in the limit of infinitesimal net-spin polarization. The prefactor of $4\pi/\hbar$  instead of $2\pi/\hbar$ denotes the fact that the net number of spin-polarized electrons changes by two with each spin flip. It is noted that only first-order processes are relevant for calculation of $M_{sf}(\mathbf{k}_i,\mathbf{k}_f)$; we have found zero net contribution from second-order processes in which an electron undergoes intra-axis scattering via two virtual elementary inter-axis scattering events.

We consider three types of  spin-flip processes due to intra-axis scattering. Two of the three are Elliott processes in which the spin flip is governed by the spin-orbit coupling of the host crystal, whereas the scattering potential is spin independent.\cite{Elliott_PR54a,Pikus_OO84} One Elliott process involves long-range interaction with the ionized-impurity potential, and the second one is a central cell correction coming from short-range interaction with the spin-independent part of the impurity potential. Using the  $\mathbf{k}$$\cdot$$\mathbf{p}$ expansion in (\ref{eq:a1}), an Elliott spin-flip matrix element has the form,  
\begin{subequations} \label{eq:elliott} 
\begin{eqnarray}
        M_{sf}^{E,i}(\mathbf{k}_i,\mathbf{k}_f) \!\!\!&\simeq& \!\!\! \mathbf{A}^T\!(\mathbf{k}_f) \mathcal{H}_E \mathbf{B}(\mathbf{k}_i) \!-\! \mathbf{B}^T\!(\mathbf{k}_f) \mathcal{H}_E \mathbf{A}(\mathbf{k}_i),\,\,\,\,\,\,\,\,\,\,\,\, \label{eq:elliott_intra} \\
       M_{sf}^{E,g}(\mathbf{k}_i,\mathbf{k}_f) \!\!\!&\simeq&\!\!\!  \mathbf{A}^T\!(\mathbf{k}_f) \mathcal{H}_E \tilde{\mathbf{B}}(\mathbf{k}_i) \!-\! \mathbf{B}^T\!(\mathbf{k}_f) \mathcal{H}_E \tilde{\mathbf{A}}(\mathbf{k}_i),\,\,\,\,\,\,\,\,\,\,\,\,  \label{eq:elliott_g}  
\end{eqnarray}
\end{subequations}
for the intravalley and intrevalley \textit{g}-process, respectively. $\mathcal{H}_E$ is a $4\times4$ interaction matrix whose form will be deduced from the symmetry of the spin-independent potential. Following (\ref{eq:8}) and (\ref{eq:5}), $\tilde{\mathbf{A}}$ and $\tilde{\mathbf{B}}$ in the $g$-process matrix element are found by exchanging the coefficients of $X_1^{2'}$ and $X_1^1$ as well as of $X_4^{x}$ and $X_4^y$ in (\ref{eq:a1}).

The last intra-axis scattering that we will consider in this work is a Yafet process in which the spin flip is governed by the spin-orbit coupling of the scattering potential.\cite{Yafet_SSP63,Pikus_OO84} Although much weaker than the dominant inter-axis Yafet mechanism, both originate from short-range interaction with the spin-orbit coupling of the impurity. The spin-flip matrix elements have the form,
\begin{subequations} \label{eq:yafet}
\begin{eqnarray}  
M_{sf}^{Y,i}(\mathbf{k}_i,\mathbf{k}_f) &\simeq&  \mathbf{A}^T(\mathbf{k}_f) \mathcal{H}_Y \mathbf{A}(\mathbf{k}_i)\,, \label{eq:yafet_intra}   \\
M_{sf}^{Y,g}(\mathbf{k}_i,\mathbf{k}_f) &\simeq&  \mathbf{A}^T(\mathbf{k}_f) \mathcal{H}_Y \tilde{\mathbf{A}}(\mathbf{k}_i)  \label{eq:yafet_g} \,, 
\end{eqnarray}
\end{subequations}
where we have neglected the contribution from the cross products of $\mathbf{B}$ vectors due to the smallness of the spin-orbit coupling in silicon. It is noted that Elliott and Yafet matrix elements can become comparable if the smallness of the nonzero elements in $\mathbf{B}$  (compared with those in $\mathbf{A}$) is compensated by the smallness of the elements in $\mathcal{H}_Y$ (compared with those $\mathcal{H}_E$). In fact, this scenario applies in the case of the electron-phonon interaction for which the spin-orbit coupling of the host atoms drives both the terms in $\mathbf{B}$ and $\mathcal{H}_Y$.\cite{Song_PRB12}

\subsection{Long-range Coulomb potential (Elliott)}\label{sec:CoulElliot}

We first consider electron scattering off  the long-range Coulomb potential of ionized donor impurities, 
\begin{equation}  \label{eq:col_simple} 
V(\mathbf{r}) = \frac{4\pi e^2}{\epsilon r} e^{-\kappa r}\,\,\,, \qquad \kappa = \sqrt{\frac{4\pi e^2 N_D}{\epsilon k_BT}}\,\,,
\end{equation}
where $\kappa$ is the Thomas-Fermi screening wavenumber, $e$ is the electron charge, and $\epsilon$ is the dielectric constant. The long-range nature of this scattering stems from the fact that the screened Coulomb potential decays on a much longer length scale compared with the lattice constant,  $\kappa a \ll 1$. We note that while scattering of this potential dominates momentum relaxation in highly doped silicon,\cite{Ridley_Book} its role in the context of spin relaxation is marginal compared with the inter-axis short-range scattering in unstrained silicon.\cite{Song_PRL14} The long-range and radial symmetry of the screened potential  allows us to consider $\mathcal{H}_E$ as a product between a unity matrix and the Fourier transform of the screened coulomb potential, 
\begin{equation}  \label{eq:65} 
 \mathcal{H}_E = V_{s}(\mathbf{q}) \mathds{1}\,\,\,,\qquad V_s(\mathbf{q}) = \frac{4\pi e^2}{\epsilon V (q^2 + \kappa^2)}\,\,,
\end{equation}
where $\mathbf{q}=\mathbf{k}_i-\mathbf{k}_f$. Substituting (\ref{eq:65}) and (\ref{eq:a1}) in (\ref{eq:elliott_intra}), the long-range intravalley spin-flip matrix element reads 
\begin{subequations} \label{eq:elliott_LR}
\begin{equation}  \label{eq:M_Ei} 
M_{E,i}^{lr}(\mathbf{k}_i,\mathbf{k}_f)  = \frac{\Delta '_X P}{E_g^2} (k_{+,i} -  k_{+,f} ) V_s(\mathbf{q})\,\,\,, 
\end{equation}
where $k_{+,i(f)} = k_{x,i(f)} + i k_{y,i(f)}$. Similarly, the spin-flip matrix element for the intervalley $g$-process is found by substituting (\ref{eq:65}) and (\ref{eq:a1}) in (\ref{eq:elliott_g}), 
\begin{equation}  \label{eq:M_Eg} 
M_{E,g}^{lr}(\mathbf{k}_i,\mathbf{k}_f)  = \frac{\Delta '_X P}{E_g^2} (k_{-,i} +  k_{-,f} ) V_s(\mathbf{q}_g + \mathbf{q})\,\,\,, 
\end{equation}
\end{subequations}
where $k_{-,i(f)} = k_{x,i(f)} - i k_{y,i(f)}$ and $\mathbf{q}_g = 0.3\mathbf{k}_X$. The long-range nature of the Coulomb potential renders the intravalley process much stronger, $V_s(\mathbf{q}\rightarrow 0) \gg  V_s(\mathbf{q}_g) $. 

\subsection{Spin-orbit coupling of impurities (Yafet)}\label{sec:yafet-spin-flip}

The second spin-flip process we consider is due to the spin-orbit coupling of the donor impurities. Their presence lowers the diamond point-group symmetry from $\hat I\otimes T_d$ to~$T_d$. Inspecting the symmetry operations of the space group $G^2_{32}$, we identify $M_2'$ as the IR that can represent the lowered symmetry of the impurity potential. Compared with the identity IR ($M_1$) whose characters are all `1',  the characters of $M_2'$ are negated for all symmetry operations that involve exchanging the two atoms in the unit cell. As elaborated on in Appendix~\ref{sec:extrinsic}, the form of  $\mathcal{H}_Y$ in (\ref{eq:yafet}) is extracted from the following considerations. Firstly, we identify the selection rules of $M_2'$ with IRs whose transformation properties match those of  transverse vector components (such as $k_x$ and $k_y$) and transverse pseudovector components (such as $\sigma_x$ and $\sigma_y$). These IRs are represented by $M_5$ and $M_5'$, respectively, where the selection rules follow
\begin{equation}\label{eq:selRules} 
M_2' \otimes  M_5 = M_5' \,\,\,,\qquad M_2' \otimes  M_5' = M_5 \,.
\end{equation}
That is, coupling to the impurities transforms a vector-type interaction to a pseudovector one and vice versa.  Next, we use these selection rules to construct $\mathcal{H}_Y$ due to the presence of impurities. Specifically, we look for terms that stem from the coupling between valence and conduction states since this coupling corresponds to transverse vector and pseudovector terms in the Hamiltonian ($X_1 \otimes X_4 = M_5 \oplus M_5'$). For the Yafet process, $\mathcal{H}_Y$ is constructed by replacing the $\mathbf{k}_{x,y}$ terms with $\sigma_{x,y}$, alongside replacement of crystal parameters with impurity ones (e.g., $Pk_i  \rightarrow \delta_B \Delta_{so}$).  Using this procedure, the leading Yafet term has the form (Appendix~\ref{sec:extrinsic}),
\begin{equation}\label{Yafet_Mat} 
\mathcal{H}_Y =  -i\delta_{B} \Delta_{so}(\rho _y\otimes\sigma _x+i\rho _z\otimes\sigma _y) \,.
\end{equation}
Substituting (\ref{Yafet_Mat}) and (\ref{eq:a1}) in (\ref{eq:yafet}), the intravalley and $g$-process spin-flip matrix elements  follow 
\begin{subequations} \label{eq:yafet_SR}
\begin{eqnarray}
M_{Y,i}(\mathbf{k}_i,\mathbf{k}_f)   &=& \frac{\delta_B \Delta_{so} P}{E_g}  (k_{-,i} +  k_{-,f} ) \,\,, \label{M_Yi}  \\
M_{Y,g}(\mathbf{k}_i,\mathbf{k}_f)  &=& i\frac{\delta_B \Delta_{so} P}{E_g}  (k_{+,i} +  k_{+,f} ) \,\, \label{M_Yg} . 
\end{eqnarray}
\end{subequations}

\subsection{Central cell Elliot spin flip}\label{sec:ccElliot}

The final spin-flip process we consider is governed by the spin-orbit coupling of the host crystal (silicon atoms), and it takes place when electrons are scattered off the short-range and spin-independent part of the impurity potential. To derive the form of the resulting Elliott matrix, $\mathcal{H}_E$, we inspect the coupling within conduction-band basis states (the conduction-valence coupling gives rise to the Yafet process as discussed in the previous section). The selection rule for coupling between conduction states follows from 
\begin{equation}\label{eq:selRules_X1X1} 
X_1 \otimes  X_1   =  M_1 \oplus M_2'   \oplus M_3' \oplus M_4 \,.
\end{equation}
Relevant to our study are $M_2'$ and the identity IR $M_1$. The identity IR represents the radial part of the central cell correction, and as such it gives rise to diagonal coupling between $X_1^1$ states or between $X_1^{2'}$  states. On the other hand, $M_2'$ represents the lowered symmetry part of the impurity potential. Its transformation properties gives rise to off-diagonal coupling between $X_1^1$ and $X_1^{2'}$ states.\cite{Song_PRB12} We therefore have two terms in the short-range Elliott matrix,
\begin{equation}\label{Elliott_SR_Mat} 
\delta H_c = \delta_B \Delta_{0}\rho_0 \otimes\sigma _0 + \delta_B \Delta_{1}\rho _y\otimes\sigma _0.
\end{equation}
where $\sigma_0$ is a $2\times 2$ unity matrix acting in spinor space,
and $\Delta_{0}$ and $\Delta_{1}$ are the diagonal and off-diagonal scattering constants. To estimate their values, we make use of the fact that the short-range and spin-independent part of the impurity potential splits the sixfold degenerate ground state energy due to valley-orbit coupling.\cite{Kohn_SSP57} Similar to the Yafet process for which the scattering amplitude was estimated from $\Delta_{so}$ (the spin-splitting of $T_2$ due to the spin-orbit coupling of the impurity), the values of $\Delta_{0}$ and $\Delta_{1}$ can be uniquely determined via the empirically known spin-independent binding energies of the ground state ($A_1$, $E$ and $T$). Below we use $\Delta_{0}\simeq 4$~meV and $\Delta_{1} \simeq 1.5$~meV, following the work of  Friesen who studied the Stark Effect for donors in silicon.\cite{Friesen_PRL05} Contrary to $\Delta_{so}$, the values of $\Delta_{0}$ and $\Delta_{1} $ are largely insensitive to the identity of the substitutional donor (Sb, As, and P).\cite{Ramdas_RPP81,Castner_PR63}

Substituting (\ref{Elliott_SR_Mat}) and (\ref{eq:a1}) in (\ref{eq:elliott}), the short-range intravalley and $g$-process spin-flip matrix elements  follow 
\begin{subequations} \label{eq:elliott_SR}
\begin{eqnarray} 
M_{E,i}^{sr}(\mathbf{k}_i,\mathbf{k}_f)   &=&  i\frac{\delta_B \Delta_{1} \eta}{4\Delta_c}  (k_{+,i} +  k_{+,f} ) \,\,, \label{M_E_SR_i}  \\
M_{E,g}^{sr}(\mathbf{k}_i,\mathbf{k}_f)  &=& -\frac{\delta_B \Delta_{0} \eta}{4\Delta_c}  (k_{+,i} +  k_{+,f} ) \,\,  \label{M_E_SR_g} . 
\end{eqnarray}
\end{subequations}

\subsection{Total intra-axis spin-flip matrix elements}\label{sec:totalM}
The total matrix element is denoted by the sum of (\ref{eq:elliott_LR}), (\ref{eq:yafet_SR}), and (\ref{eq:elliott_SR}). For spin flips due to intravalley and $g$-process, one gets 
\begin{subequations} \label{eq:M}
\begin{eqnarray}  
\!\!\! M_{i}(\mathbf{k}_i,\mathbf{k}_f)  \!\!\! &\simeq&\!\!\!  \frac{P}{E_g} \delta_B \Delta_{so} \left( k_- + C_{\mathbf{q}}q_+ + D_1  k_+\right)\!,\,\,\,\,\,\,\,\,\,\,\,\,\,\, \label{eq:M_i}  \\
\!\!\! M_{g}(\mathbf{k}_i,\mathbf{k}_f) \!\!\! &\simeq& \!\!\! \frac{iP}{E_g} \delta_B \Delta_{so} \left( k_- + C_{\mathbf{q}_g}q_- -  D_0  k_+\right)\!,\,\,\,\,\,\,\,\,\,\, \label{eq:M_g} 
\end{eqnarray}
where $k_{\pm} = k_{\pm,i} +  k_{\pm,f}$, $q_{\pm} = k_{\pm,i} -  k_{\pm,f}$, and 
\begin{eqnarray} \label{eq:M_i}
C_{\mathbf{p}} = \frac{V_s(\mathbf{p})\Delta '_X}{\delta_BE_g \Delta_{so}}\,\,,\qquad   D_{1(0)} = \frac{2\Delta_{1(0)}\Delta_X}{\Delta_c \Delta_{so}}\,.
\end{eqnarray}
\end{subequations} 
The $C$ and $D$ terms represent the long- and short-range Elliott processes, respectively. For Si:Sb in which the spin-orbit coupling of the impurity is relatively strong ($\Delta_{so} \simeq 0.3$~meV),\cite{Aggarwal_PR65,Castner_PR67} both Elliott processes can be neglected and the spin relaxation is governed by the Yafet process  ($C,D \ll 1$).  Only for the case of Si:P ($\Delta_{so} \simeq 0.03$~meV),\cite{Castner_PR67}  the Elliott processes become comparable to the Yafet one. This result is not surprising given that the spin-orbit coupling of silicon is smaller than that of antimony and arsenic while being comparable to that of phosphorus. In addition, we can quantify the ratio between intra-axis and inter-axis spin-flip matrix elements in unstrained silicon by comparing (\ref{eq:M}) and (\ref{eq:M_appr}). Using the facts that in semiconductors $P/E_g \sim a/2\pi$ and $k^2 \sim 2mk_BT/\hbar^2$, the ratio is of the order of $2mk_BTa^2/h^2$ revealing that intra-axis spin-flip matrix element is about three orders of magnitude weaker than the inter-axis one at room temperature (and more than that at lower temperatures).   

Finally, we note on the qualitative  difference between spin flips caused by electron-phonon and electron-impurity interactions.  In the case of phonons, the spin-orbit coupling of the host crystal drives both the Elliott and Yafet processes, giving rise to cancellation of the leading Elliott and Yafet intravalley processes when space inversion symmetry is respected.\cite{Yafet_SSP63,Song_PRB12} Yafet found that in silicon this cancelation gives rise to quadratic rather than linear dependence of the intravalley spin-flip matrix element on the transverse components of the acoustic-phonon wavevector, ($\propto q_{\pm}^2$).\cite{Yafet_SSP63}  On the other hand,  the intra-axis spin-flip matrix elements due to scattering off impurities have linear dependence on the transverse crystal momentum of the initial and final states, as shown in (\ref{eq:M}).  The reason for the linear dependence is that there is no Elliott-Yafet cancellation when dealing with impurities, whose presence breaks the space inversion symmetry and whose spin-orbit coupling is not related to the one of the host-crystal atoms.  Quantitatively, this difference can be seen from the lack of interference terms between the short-range Yafet, short-range Elliott,  and long-range Elliott processes after averaging the value of $|M_{i/g}(\mathbf{k}_i,\mathbf{k}_f)|^2$ over angular angles.

\section{Anisotropy and Overall spin relaxation in strained silicon} \label{sec:relax}

So far, we have assumed spin orientation along the valley axis, $\mathbf{s} \parallel \mathbf{z}$. The matrix elements for arbitrarily orientation, $\mathbf{s} \nparallel \mathbf{z}$, can be related with those in (\ref{eq:M})   by\cite{Song_PRB12}
\begin{eqnarray}  \label{eq:spinOrientation}  
    \langle \Uparrow_{\mathbf{s}} \mathbf{k}_f | V |\Downarrow_{\mathbf{s}}  \mathbf{k}_i\rangle  &=&    \cos^2\frac{\theta}2e^{i\phi} \langle \Uparrow_{\mathbf{z}}  \mathbf{k}_f | V |\Downarrow_{\mathbf{z}} \mathbf{k}_i \rangle  \,\,\,\,\,\,\,\,\,\,\, \\ && -    \sin^2\frac{\theta}2e^{-i\phi} \langle \Downarrow_{\mathbf{z}} \mathbf{k}_f | V |\Uparrow_{\mathbf{z}} \mathbf{k}_i\rangle  \,,  \nonumber
\end{eqnarray}
where $\theta$ is measured from the valley axis and $\phi$ is the azimuthal angle. We have omitted terms proportional to $\langle \Downarrow_z \mathbf{k}_f | V |\Downarrow _z \mathbf{k}_i\rangle  - \langle \Uparrow_z \mathbf{k}_f | V |\Uparrow _z \mathbf{k}_i\rangle $ since they vanish identically for linear in $\mathbf{k}$ terms (\ref{eq:M}). Similar to the case of phonon-induced spin relaxation,\cite{Song_PRB12} summing the contributions from all six valleys in unstrained silicon yields isotropic spin relaxation. The anisotropy emerges when applying strain, yielding as twice as strong intra-axis spin relaxation when the low-energy valleys pair and  spin orientation are collinear compared with the perpendicular case.

The total spin relaxation rate is the sum of the intervalley $f$-process (inter-axis scattering), intravalley and intervalley $g$-process (intra-axis scattering),
\begin{eqnarray}  \label{total_tau_f_i_g}
\frac{1}{\tau_{s}} =  \frac{1}{\tau_{s}^{\text{inter}}}  +  \frac{1}{\tau_{s}^{\text{intra}}}   \,\,.\,\,\,\,\,\,\,\,\,\, 
\end{eqnarray}
The effect of compressive strain on the intra-axis and inter-axis processes is different since only the latter can be completely quenched  when the valley splitting energy is large: $1/\tau_{s}^{\text{inter}} \rightarrow 0$ when $e^{-\Delta_v/k_BT} \ll 1$, where $\Delta_v$ is the strain-induced valley splitting energy.  On the other hand, the intra-axis relaxation rate, $1/\tau_{s}^{\text{intra}}$, is only mildly affected via the emerged anisotropy. Below we focus on the two extreme cases for parallel and perpendicular orientations of strain and spin axes, and express the total spin relaxation in silicon due to electron-impurity and electron-phonon as functions of temperature, donor concentration, and valley splitting energy (for uniaxial compressive strain along one of the equivalent [001]-crystallographic directions).

\subsection{Total inter-axis spin relaxation ($f$-process)}\label{sec:total_f}
\begin{figure}
\includegraphics[width=8.6cm]{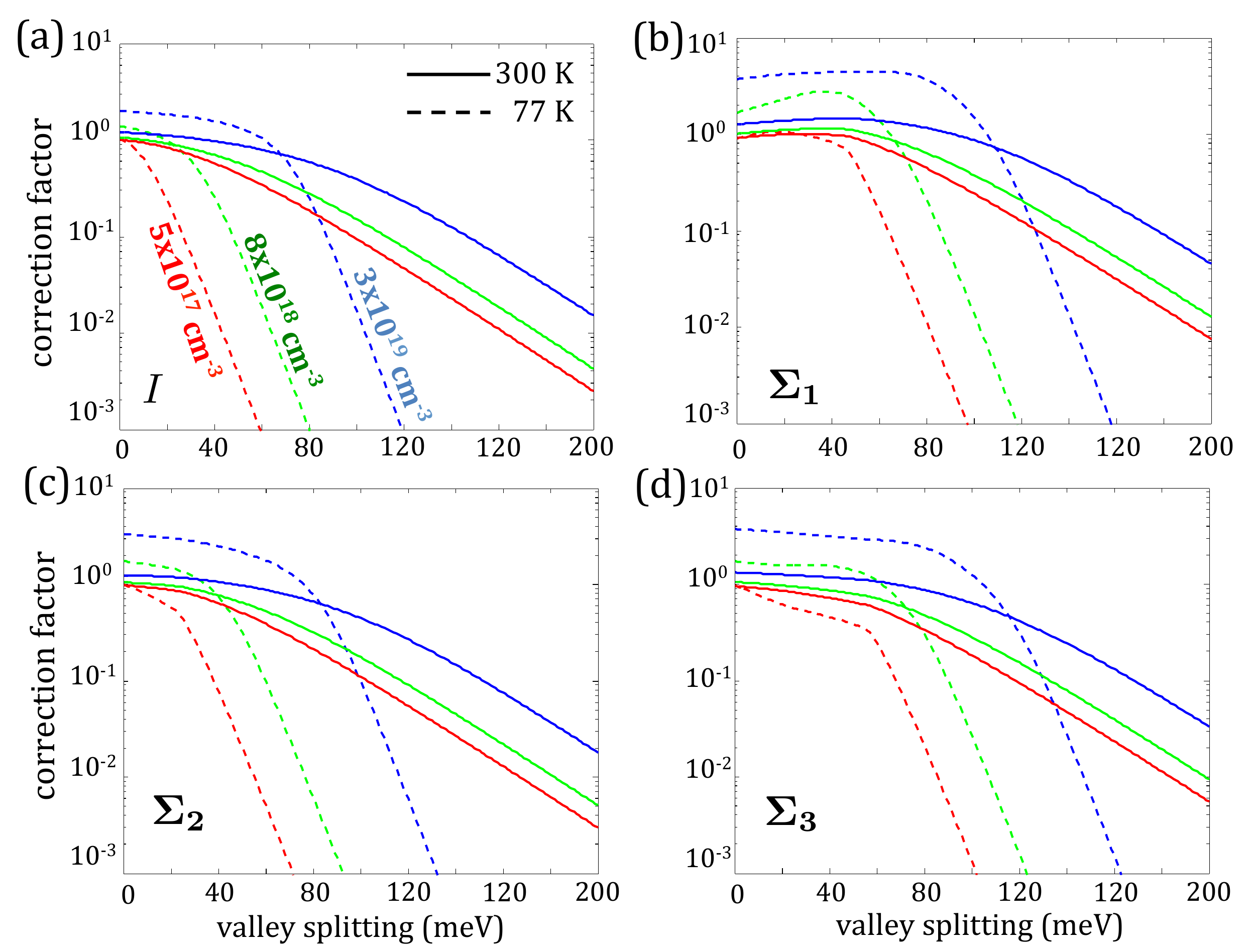}
\caption{Correction factors of the intervalley $f$-process as a function of the strain-induced valley splitting energy. The solid (dashed) lines show room temperature (77~K) results for three different densities ($5\times10^{17}$, $8\times10^{18}$, and $3\times10^{19}$ cm$^{-3}$).  The electron-impurity correction factor is shown in (a), and the electron-phonon correction factors are shown in (b)-(d) for each of the three symmetry-allowed phonons modes ($\Sigma_{1-3}$). All of the results are shown for parallel configuration between the strain and spin orientation axes. The results for the perpendicular configuration are similar in nature.  \label{fig:cf}}
\end{figure}

The $f$-process spin relaxation is expressed by,
\begin{eqnarray}  \label{tau_inter_axis_strain}
\frac{1}{\tau_{s}^{\text{inter}}}    =  \frac{C_{I,f}}{\tau_{I,f}} + \sum_{j=1,2,3} \frac{C_{\Sigma_j}}{\tau_{\Sigma_j}}
\end{eqnarray}
where the first term denotes the contribution due to scattering off impurities,\cite{Song_PRL14} and the sum denotes the contribution from the three $\Sigma$-axis  phonon modes that govern the spin-flip $f$-process intervalley scattering.\cite{Song_PRB12} The $\tau$ terms on the right-hand side denote the corresponding strain-free spin relaxation when assuming electron Boltzmann distribution in (\ref{eq:tau_s}). $C_{I,f}$ and $C_{\Sigma_j}$  denote correction factors caused by the strain suppression of the $f$-process (both approach zero when $e^{-\Delta_v/k_BT} \ll 1$), as well as the correction to the spin relaxation when deviating from the Boltzmann limit (high density and low temperature). The electron-impurity spin relaxation rate constant in (\ref{tau_inter_axis_strain}) follows,
\begin{eqnarray}  \label{tau_inter_imp_Boltzmann}
\frac{1}{\tau_{I,f}} = \frac{N_Da_B^3}{\tau_{D}} \sqrt{\frac{T}{T_R}}, \qquad  \frac{1}{\tau_{D}} = \frac{16\sqrt{\pi}}{3\hbar} \frac{\Delta_{so}^2}{E_R}
\end{eqnarray}
$E_R = k_BT_R = \hbar^2/2m_d a_B^2 \sim 35$~meV is the effective Rydberg energy in silicon, where $m_d=0.32m_0$ is the effective density-of-state mass and $a_B \simeq 1.85$~nm. The value of $\tau_D$ is about 30~ps for Sb, 240~ps for As, and 3~ns for P. The electron-impurity correction factor in (\ref{tau_inter_axis_strain}) is expressed by,
\begin{eqnarray}  \label{tau_inter_imp_F}
C_{I,f}  = \frac{3}{4}\frac{ (1 + \delta_{s,v} )\mathcal{I}_f(\Delta_v,\Delta_v) +(3 - \delta_{s,v} )\mathcal{I}_f( \Delta_v,0)}{\mathcal{I}_{1}(0)+2\mathcal{I}_{1}( \Delta_v)}.\,\,\,\,
\end{eqnarray}
$\delta_{s,v}=1 (0)$  if the spin orientation is collinear (perpendicular) to the low-energy valley axis. The other terms are defined by the two integrals,
\begin{eqnarray}  \label{tau_inter_imp_If}
\mathcal{I}_f(\varepsilon_1,\varepsilon_2) = \int_{\varepsilon_m}^\infty \! d\varepsilon \frac{\sqrt{(\varepsilon-\varepsilon_1)(\varepsilon-\varepsilon_2)}}{k_BT}     \frac{\partial \mathcal{F}}{\partial \varepsilon}\,, \,\,\,\,\,\,\,\,\,
\end{eqnarray}
where $\varepsilon_m = \text{max}\{\varepsilon_1,\varepsilon_2\}$, and 
\begin{eqnarray}  \label{tau_inter_imp_IN}
\mathcal{I}_{\,n}(\varepsilon_1) = \frac{2}{\sqrt{\pi}} \int_{\varepsilon_1}^\infty \! d\varepsilon \left(\frac{\varepsilon-\varepsilon_1}{k_BT}\right)^{\!\!n/2}  \frac{\partial \mathcal{F}}{\partial \varepsilon}.
\end{eqnarray}
In the following, we will show results when using the Fermi-Dirac distribution to represent $\mathcal{F}(\varepsilon)$.  In the Boltzmann limit where $\text{max}\{\mathcal{F}(\varepsilon)\} \ll 1$, both integrations can be performed analytically yielding $\mathcal{I}_{1}(\varepsilon_1) \rightarrow e^{-\varepsilon_1/k_BT}$ and $\mathcal{I}_f(\varepsilon_1,\varepsilon_2)  \rightarrow r_d e^{-r_a} K_1( r_d)$, where $r_d = |\varepsilon_2-\varepsilon_1|/2k_BT$, $r_a = (\varepsilon_2+\varepsilon_1)/2k_BT$,  and  $\text{K}_1$ is the first-order modified Bessel function of the second kind. Figure~\ref{fig:cf}(a) shows the value of $C_{I,f}$ at 77~K (dashed lines) and 300~K  (solid lines) as a function of valley splitting for three donor densities. The value of $C_{I,f}$ at zero strain approaches unity at low densities.

The  $f$-process spin relaxation in  (\ref{tau_inter_axis_strain}) due to electron-phonon interaction is decomposed in a similar way. The rate constants denote the corresponding strain-free spin relaxation in the Boltzmann limit,\cite{Song_PRB12}
\begin{eqnarray}  \label{tau_inter_phonon_Boltzmann}
\frac{1}{\tau_{\Sigma_j}} =  \frac{\sqrt{r_j}}{\tau_{j}}   \frac{K_1(r_j)}{\text{sinh}(r_j)}.
\end{eqnarray}
The temperature dependence is carried by the parameter $r_j = T_j/2T$, where $T_1=540$~K, $T_2=660$~K, and $T_3=270$~K are the energies of the three types of symmetry-allowed shortwave phonon modes, $\Sigma_{1-3}$. The time constants $\tau_{j}$ are governed by the corresponding electron-phonon spin-flip matrix elements,\cite{Song_PRB12} yielding $\tau_{1} \sim 20$~ns,  $\tau_{2} \sim 70$~ns, and $\tau_{3} \sim 200$~ns. The expressions for the phonon-related strain suppression factors are cumbersome ($C_{\Sigma_j}$ in (\ref{tau_inter_axis_strain})), and we present them in Appendix~\ref{App:Fterms}. Their dependence on the strain-induced valley splitting energy is shown in Figs.~\ref{fig:cf}(b)-(d).

\subsection{Total intra-axis spin relaxation}\label{sec:total_intra_axis}
The spin relaxation rate due to intravalley and intervalley $g$-processes in (\ref{total_tau_f_i_g})  has four contributions,
\begin{eqnarray}  \label{tau_intra_axis_strain}
\frac{1}{\tau_{s}^{\text{intra}}}    =  \frac{C_{I,\text{sr}}}{\tau_{I,\text{sr}}} +  \frac{C_{I,\text{lr}}}{\tau_{I,\text{lr}}} + \frac{C_{\text{ac}}}{\tau_{\text{ac}}} +  \frac{C_{\Delta}}{\tau_{\Delta}}\,.
\end{eqnarray}
The $\tau$ terms on the right-hand side are spin relaxation times when assuming electron Boltzmann distribution in unstrained silicon. The $C$ terms are corrections due to the strain and/or deviations from Boltzmann statistics. The first term on the right-hand side denotes the contribution from electron scattering off the short-range impurity potential. Qualitatively, this mechanism influences the intravalley and intervalley $g$-process similarly. The second term denotes the contribution from electron scattering off the ionized impurity potential. As mentioned at the end of Sec.~\ref{sec:CoulElliot}, the effect of this long-range scattering potential on the intervalley $g$-process is negligible compared with intravalley one. The third and fourth terms  denote contributions from electron interaction with long-wavelength acoustic phonons (intravalley) and $\Delta$-axis shortwave phonons (intervalley $g$-process). We recall that a major difference from the inter-axis case is that  the intra-axis relaxation does not change appreciably when the valley-splitting energy is large. Here, the intra-axis $C$ terms only bring out the anisotropy in spin relaxation; they do not become negligible at large strain levels.

\begin{figure*}
\includegraphics[width=16cm]{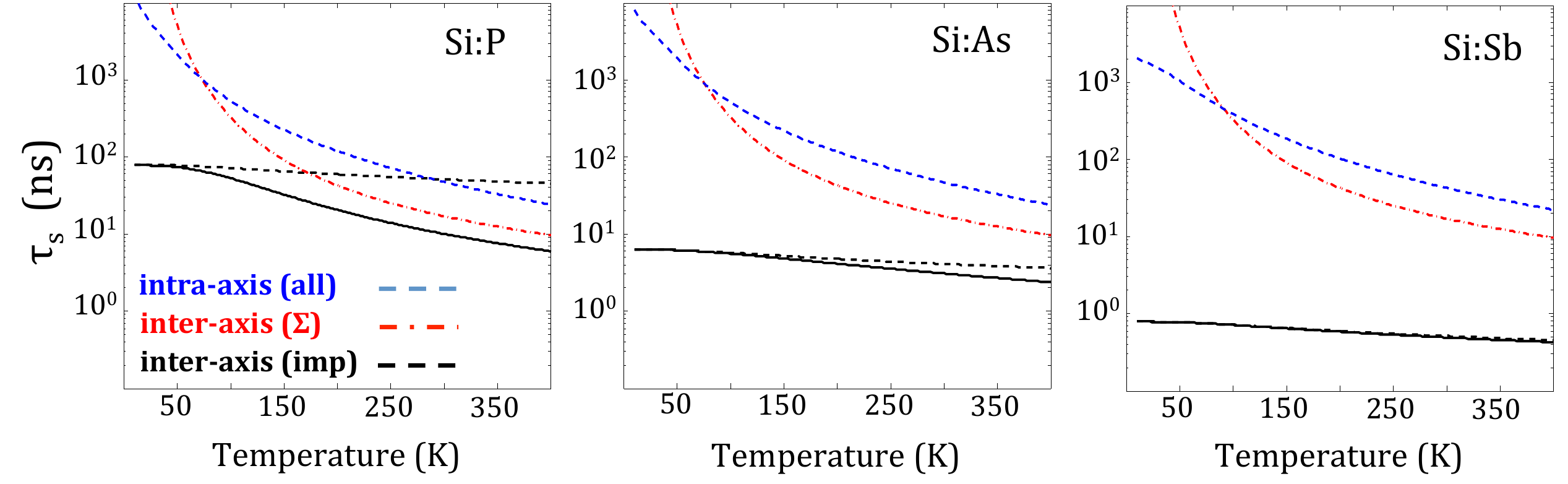}
\caption{Spin relaxation in unstrained $n$-type silicon doped with P (left), As (middle), and Sb (right) as a function of temperature. The donor concentration is $N_D$=10$^{19}$~cm$^{-3}$. The solid black lines denote the total relaxation time. The inter-axis mechanisms from phonons and impurities are shown by the dotted-dashed red lines and dashed black lines, respectively. The intra-axis mechanism is shown by the dashed blue line, and it includes contributions from both phonons and impurities.  \label{f:T}}
\end{figure*}

Starting with the intra-axis electron-impurity processes, we calculate the resulting spin relaxation by substituting (\ref{eq:M}) into (\ref{eq:tau_s}).  For the short-range interaction, denoted by the first term in  (\ref{tau_intra_axis_strain}), we get that the spin relaxation in the Boltzmann limit yields
\begin{eqnarray}  \label{tau_intra_imp_Boltzmann}
\frac{1}{\tau_{I,\text{sr}}} = \alpha_D \frac{N_Da_B^3}{\tau_{D}} \left(\frac{T}{T_R}\right)^{\!\frac{3}{2}}\!\!, \,\,  \alpha_D \!= \! \frac{(4+2D_0^2+2D_1^2)P^2}{3\pi^2 a_B^2 E_g^2},\,\,\,\,
\end{eqnarray}
where $\tau_D$ and $T_R$ were defined in (\ref{tau_inter_imp_Boltzmann}), and $D_{0,1}$ in (\ref{eq:M_i}). Compared with $\tau_D$, $\alpha_D$ has a weaker dependence on donor identity, ranging from 1.6$\times$10$^{-3}$ for Si:Sb and 5.7$\times$10$^{-3}$ for Si:P. Compared with the impurity-induced inter-axis spin relaxation in  (\ref{tau_inter_imp_Boltzmann}), the smallness of $\alpha_D$  demonstrates the negligible effect  of the intra-axis impurity scattering on spin relaxation in unstrained silicon.  When deviating from the Boltzmann limit, the strain-dependent correction factor in (\ref{tau_intra_axis_strain}) follows
\begin{eqnarray}  \label{C_intra_sr_imp}
C_{I,\text{sr}}  = \frac{3\sqrt{\pi}}{16}\frac{ (1 + \delta_{s,v} )\mathcal{I}_4(0) +(3 - \delta_{s,v} )\mathcal{I}_4( \Delta_v)}{\mathcal{I}_1(0)+2\mathcal{I}_1( \Delta_v)}\,,\,\,\,
\end{eqnarray}
where $\mathcal{I}_n(\varepsilon)$ was defined in (\ref{tau_inter_imp_IN}). Turning to the long-range interaction, denoted by the second term in  (\ref{tau_intra_axis_strain}), we get that the spin relaxation in unstrained silicon in the Boltzmann limit yields 
\begin{subequations} \label{eq:lr}
\begin{eqnarray}  \label{tau_intra_imp_lr_Boltzmann}
\frac{1}{\tau_{I,\text{lr}}} =  \frac{N_Da_B^3}{\tau_{\text{lr},0}} \sqrt{\frac{T_R}{T}} \mathcal{L}(r_N), \,\,\,\,
\end{eqnarray}
where $r_{N} = N_D/N_T$ and
\begin{eqnarray}  \label{eq:beta}
\tau_{\text{lr},0}          &=&  \frac{9\hbar a_B^2 E_g^4}{128\sqrt{\pi}P^2E_R \Delta_X^2} \approx 75~\text{ns},\\
 N_T &=& \frac{2 \epsilon m_d (k_BT)^2}{\pi e^2 \hbar^2} \simeq 1.45 \! \cdot \!10^{19} \!\left(\frac{T}{300}\right)^2 \!\text{cm}^{-3}, \,\,\,\,\,\,\,\,\,\,\,\,\,\,\,\,\\
 \mathcal{L}(x) &=&  - 1 - \left(1 + x\right)e^{x}E_i(-x) .
\end{eqnarray}
\end{subequations}
Here, $E_i(x)$ is the exponential integral special  function, $-E_i(-x) = \int_{x}^{\infty} dt e^t/t$ for $x>0$. Unlike the short-range mechanism, there is no dependence on donor identity (all donors yield similar long-range ionized Coulomb potential). Similar to the other intra-axis mechanisms, however, this mechanism has negligible contribution to the spin relaxation in unstrained silicon compared with the contribution from the inter-axis processes. In fact, it will be shown to be smaller than the intra-axis electron-phonon interaction as well as the intra-axis short-range impurity scattering in Si:Sb and Si:As. Following the notation used in (\ref{tau_intra_axis_strain}), the strain-dependent correction factor when deviating from the Boltzmann limit is 
\begin{subequations} \label{eq:C_lr}
\begin{eqnarray}  \label{C_intra_lr_imp}
C_{I,\text{lr}}  = \frac{3\sqrt{\pi}}{8\mathcal{L}(r_{N})} \frac{ (1 \!+\! \delta_{s,v} )\mathcal{I}_{\text{lr}}(0) +(3 \!-\! \delta_{s,v} )\mathcal{I}_{\text{lr}}( \Delta_v)}{\mathcal{I}_1(0)+2\mathcal{I}_1( \Delta_v)},\,\,\,\,\,\,
\end{eqnarray}
where  
\begin{eqnarray}  \label{integral_lr}
\mathcal{I}_{\text{lr}}(\varepsilon_0) &=&  \int_{\varepsilon_0}^\infty \! d\varepsilon  \frac{\partial \mathcal{F}}{\partial \varepsilon}  \left( \ln(1+G_{\varepsilon,\varepsilon_0}) - \frac{G_{\varepsilon,\varepsilon_0}}{1+G_{\varepsilon,\varepsilon_0}}  \right),\,\,\,\, \nonumber \\
G_{\varepsilon,\varepsilon_0} &=& \frac{1}{r_N}\frac{\varepsilon-\varepsilon_0}{k_BT} .
\end{eqnarray} 
\end{subequations}

The final spin relaxation mechanisms are those from intra-axis electron-phonon scattering, denoted by the last two terms in (\ref{tau_intra_axis_strain}). They are driven by intravalley scattering with acoustic phonons (mainly transverse modes), and $g$-process intervalley scattering with shortwave phonon modes of $\Delta_1$ symmetry.\cite{Song_PRB12} The intravalley and intervalley $g$-process strain-free rates in the Boltzmann limit follow,\cite{Li_PRL11,Song_PRB12}  
\begin{eqnarray}  \label{tau_intra_axis_phonon_Boltzmann}
\frac{1}{\tau_{\text{ac}}}  =  \frac{1}{\tau_{\text{ac},0}}  \left(\frac{T}{T_R}\right)^{\!5/2}\,,\qquad \frac{1}{\tau_{\Delta}}  =   \frac{\sqrt{r_{g}}}{\tau_{\Delta,0}}   \frac{K_2(r_g)}{\text{sinh}(r_g)},
\end{eqnarray}
where $\tau_{\text{ac},0} \sim 50$~ns and $\tau_{\Delta,0} \sim 2$~$\mu$s are governed by the respective electron-phonon spin-flip matrix elements. The temperature dependence of the $g$-process is carried by the parameter $r_{g} = T_g/2T$, where $T_g=240$~K is the energy of the relevant $\Delta_1$ shortwave phonon mode. $\text{K}_2$ is the second-order modified Bessel function of the second kind. The intravalley correction factor is similar to that of the short-range potential, $C_{\text{ac}}=C_{I,\text{sr}}$, provided in (\ref{C_intra_sr_imp}). The expression for the intervalley $g$-process correction is cumbersome ($C_{\Delta}$), and we provide it in Appendix~\ref{App:Fterms}. In the Boltzmann limit, all of the intra-axis correction factors approach
\begin{eqnarray}  \label{tau_C_intra_axis_phonon_Bolt}
\begin{array}{cc}
C_{I,\text{sr}}, & C_{I,\text{lr}},  \\
C_{\text{ac}}, & C_{\Delta}, \end{array}  \rightarrow \frac{3}{4}\left[   1 + \frac{e^{-2r_v} + \delta_{s,v}\left(1-e^{-2r_v} \right)}{1+2e^{-2r_v}} \right], \,\,\,\,
\end{eqnarray}
where $r_v = \Delta_v/2k_BT$. In this limit and for large valley-splitting energy ($e^{-2r_v} \ll 1$), the  intra-axis spin relaxation is as twice as strong when the strain and spin orientations are parallel ($\delta_{s,v}=1$) compared with the case in which they are perpendicular ($\delta_{s,v}=0$).


\section{Results and Discussion} \label{sec:result}

The analysis provided in the previous section allows us to quantify the spin relaxation time of mobile electrons in $n$-type silicon for any temperature, donor concentration, donor identity, and valley splitting energy (at the uniaxial compressive strain configuration). It takes less than a minute to generate the results of each of the figures in this work with a simple computer. 

\begin{figure*}
\includegraphics[width=16cm]{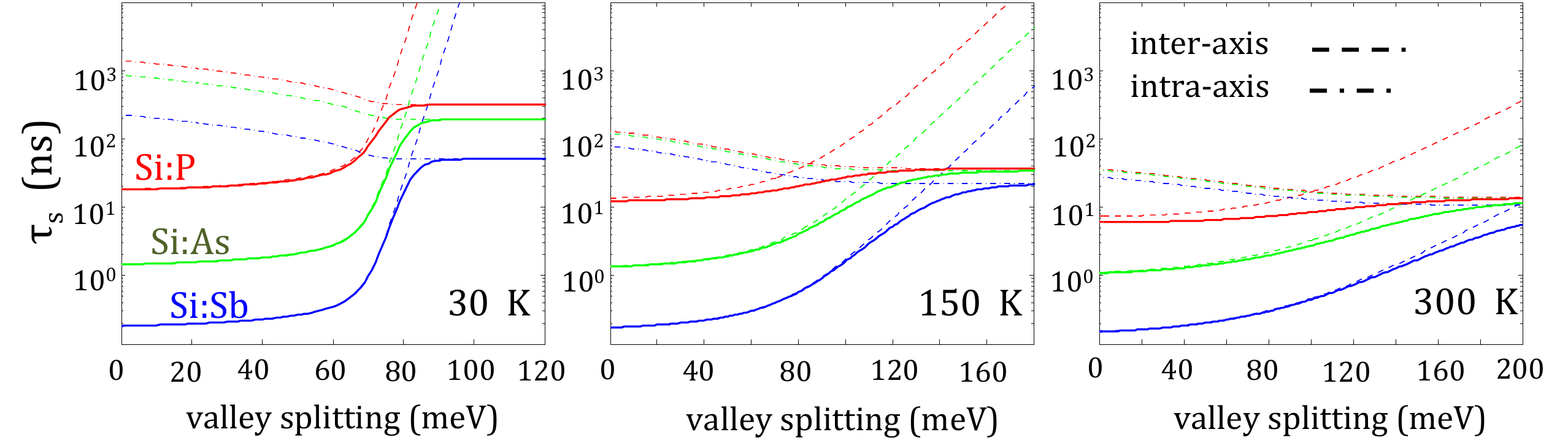}
\caption{Spin relaxation in silicon as a function of strain-induced valley splitting energy at 30~K (left), 150~K (middle), and 300~K (right).  The donor concentration is $N_D$=3$\times$10$^{19}$~cm$^{-3}$, and the strain axis is parallel to the spin orientation ($\delta_{s,v}=1$). The solid lines denote the total relaxation time. The inter-axis and intra-axis contributions are denoted by the dashed and dotted-dashed lines, respectively.  \label{f:D}}
\end{figure*}

Figure~\ref{f:T} shows the spin relaxation times in unstrained silicon as a function of temperature in Si:P, Si:As and Si:Sb. The donor concentration in all cases is $N_D$=10$^{19}$~cm$^{-3}$. For the case of Si:Sb, shown in the right panel, the relaxation is governed by inter-axis scattering off impurities at  all temperatures.\cite{Song_PRL14} In Si:P, the spin relaxation is governed by this mechanism at low temperatures and by the other inter-axis mechanism at high-temperatures (interaction with shortwave $\Sigma$-axis phonons). The intra-axis spin relaxation ($\tau_{s}^{\text{intra}}$), shown by the dashed-blue line, has marginal contribution at all cases. The small effect of the intra-axis mechanism applies at lower donor concentrations as well, in which the relaxation is largely governed by the interaction with the shortwave $\Sigma$-axis phonons.\cite{Li_PRL11} The only means to bring the intra-axis mechanism into play is by strain-induced quenching of the inter-axis mechanisms. 

\begin{figure*}
\includegraphics[width=16cm]{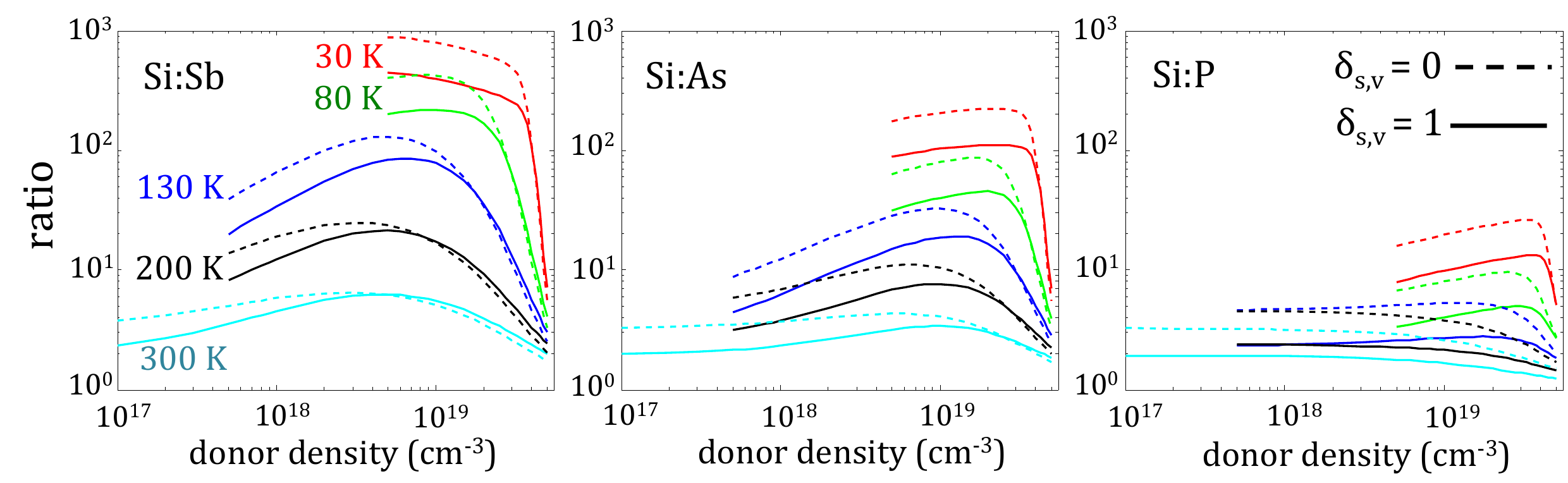}
\caption{The relative improvement in the spin relaxation time as a function of donor concentration. The $y$-axis denotes the ratio between the total spin relaxation times for $\Delta_v=100$~meV and $\Delta_v=0$~meV, where $\Delta_v$ is the strain-induced valley splitting energy. The solid (dashed) lines show the ratio when the strain and spin orientations are parallel (perpendicular) to each other. At low temperatures (30 and 80~K), the ratio is shown for densities above the metal to insulator transition ($N_D>5 \times 10^{18}$~cm$^{-3}$), and at intermediate temperatures for densities above $N_D>5 \times 10^{17}$~cm$^{-3}$ (130 and 200~K). These conditions avoid the freeze-out region in which electrons become localized by the donor potential. \label{f:r}}
\end{figure*}

Figure~\ref{f:D} shows the spin relaxation time as a function of the strain-induced valley splitting energy in Si:P, Si:As and Si:Sb at three temperatures. The donor concentration in all cases is $N_D$~=~3$\times$10$^{19}$~cm$^{-3}$. In all cases, the spin relaxation is switched from being governed by inter-axis mechanisms to intra-axis ones at large valley splitting energies. The enhancement in spin relaxation time is most evident at low temperatures because of the increased ratio between valley splitting and thermal energies as well as the weaker effect from phonon-related interactions. Furthermore, the improvement is larger than two orders of magnitude for Si:Sb at low temperatures due to quenching of its strong inter-axis impurity scattering.\cite{Song_PRL14} We also notice that while the inter-axis spin relaxation mechanisms are quenched by strain, the intra-axis spin relaxation time becomes slightly faster. The latter is explained by the increase of the chemical potential with respect to the conduction band edge. Specifically, the electron density is redistributed among six valleys at zero-strain condition while only among two valleys when the strain-induced valley splitting energy is very large. This electron redistribution leads to a change in the chemical potential which for degenerate doping such as the one studied here, $N_D$=3$\times$10$^{19}$~cm$^{-3}$, results in an increase from about 12 to 60~meV at room temperature (with respect to the conduction band edge) and from about 30~meV to 60~meV at 30~K. Given that the intra-axis spin-flip matrix elements are commensurate with the electron wavevector, the relaxation rate increase about linearly with the Fermi energy.

Figure~\ref{f:r} shows the improvement ratio in the total spin relaxation time when the strain-induced valley splitting energy is 100~meV (i.e., $\tau_s(\Delta_v=100~\text{meV})/\tau_s(\Delta_v=0~\text{meV})$).  There are several notable features. The first one is that the improvement is mostly significant at low temperatures due to the negligible population of the high-energy valleys. At room temperature, on the other hand, $\Delta_v/k_BT$ is of the order of 4, which is not sufficient to quench the inter-axis processes. Accordingly, the improvement is nearly three order of magnitude in Si:Sb at 30~K while being much smaller at room temperature. The second feature is that the ratio initially increases with donor density before sharply decaying at very large densities. The conjunction of two factors gives rise to the initial increase: (i) the electron-impurity scattering becomes more significant when increasing the donor density, and (ii) the strain quenches the inter-axis elastic scattering more effectively than the inelastic one. The latter involves $\Sigma$-axis phonons whose energy  renders the effective valley splitting smaller. Specifically, electron transitions can take place already when the electron energy is $\Delta_v-\varepsilon_{\Sigma}$ with respect to the conduction band edge  ($\varepsilon_{\Sigma}$$\,$$\sim$$\,$47~meV for the dominant $\Sigma_1$ mode). As a result, the strain quenches more effectively the elastic electron-impurity inter-axis mechanism for which electrons transitions can take place when their energy is $\Delta_v$.  This behavior explains the initial increase of the ratio in Fig.~\ref{f:r} with donor concentration (bigger role played by the electron-impurity interaction). This behavior supports the fact that the improvement is larger for Si:Sb in which the inter-axis electron-impurity scattering is strongest (compared with Si:P in which the electron-phonon interaction plays a bigger role).  Finally, the improvement in spin relaxation sharply decays for all donor types for concentrations close to  $N_D$=5$\times$10$^{19}$~cm$^{-3}$. At these densities, the chemical potential is higher than the valley splitting energy ($\mu > \Delta_v$=100~meV in this case), so that the high-energy valleys become populated and the inter-axis mechanisms are restored. In other words, the improvement in spin relaxation for degenerate doping conditions is viable when $(\Delta_v-\mu)/k_BT \gg 1$ rather than $\Delta_v/k_BT \gg 1$.

\begin{figure}
\includegraphics[width=8.6cm]{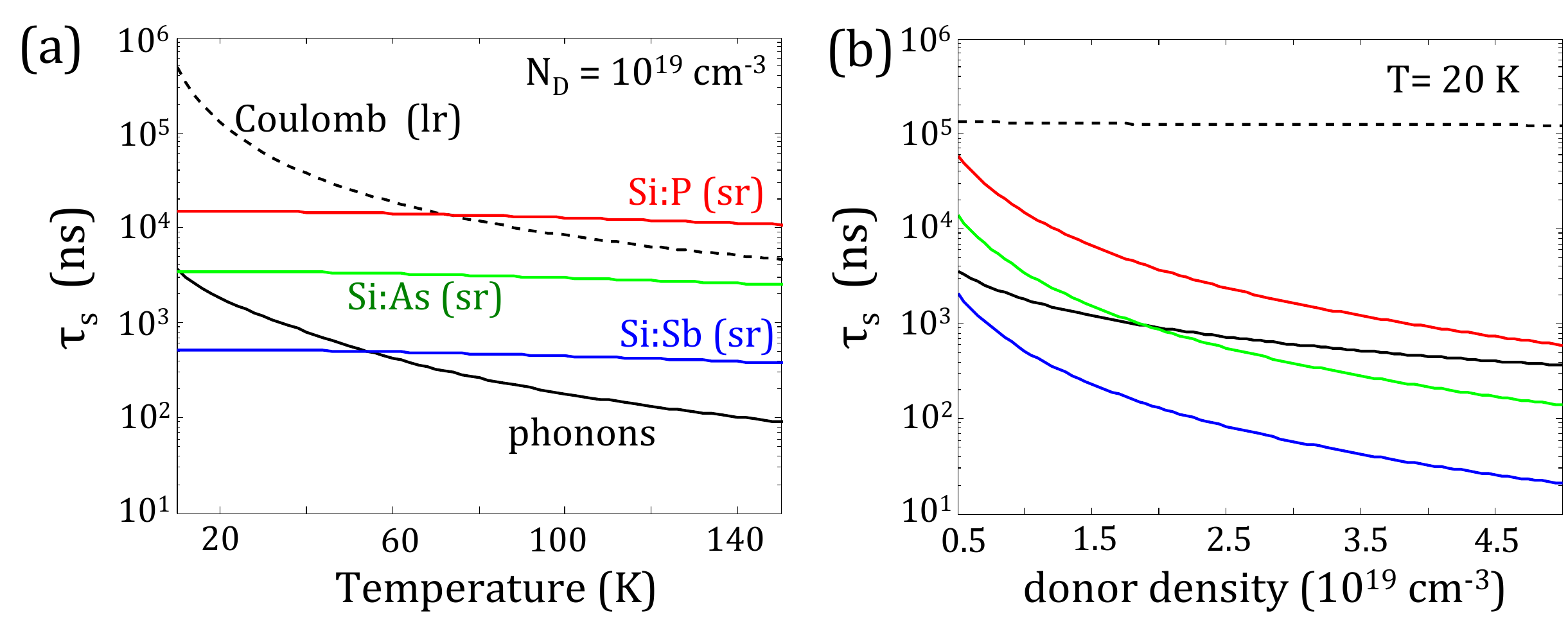}
\caption{Spin relaxation times of the various intra-axis mechanisms in strained silicon where $\Delta_v=120$~meV. (a) and (b) show the dependencies of these mechanisms on temperature when $N_D$=10$^{19}$~cm$^{-3}$ and on donor concentration when $T$=20~K, respectively.  Unlike the interaction with the short-range impurity potential, the Coulomb interaction with the long-range ionized impurity potential (dashed lines) and the interaction with phonons (black solid lines) are independent of the donor identity. \label{f:intra}}
\end{figure}

The last effect we focus on is the relative role of the intra-axis mechanisms in strained silicon. Figures~\ref{f:intra}(a) and (b) show the dependencies of these mechanisms on temperature and donor concentrations, respectively. The strain-induced spin splitting is $\Delta_v=$120~meV. We notice that the spin relaxation due to the long-range interaction of electrons with impurities is a relatively weak effect (dashed lines). This interaction shows a relatively strong dependence on temperature compared with that of the short-range interaction. The reason originates from the decreased role of screening at high temperatures; thermally agitated electrons screen less effectively and therefore the spin relaxation is enhanced. Finally, Fig.~\ref{f:intra}(b) shows an untypical trend wherein the intra-axis spin relaxation time due to phonons has a stronger dependence on donor density compared with the long-range interaction with impurities.  The weak dependence of the latter is understood by the fact that the increased donor concentration is accompanied by stronger screening, and the two effects cancel each other (i.e., the relaxation saturates).  The enhancement of the intra-axis spin relaxation due to phonons when increasing the donor concentration is understood by the rise of the chemical potential. The phonon wavevector involved in the spin-flip of the electron  is proportional to $\sqrt{E_F}$ rather than $\sqrt{k_BT}$ in degenerately doped silicon, thereby enhancing the relaxation rate with increasing the donor density.  
 
\section{Conclusion} \label{sec:conc}

We have derived the intra-axis spin-flip matrix elements due to scattering off impurities (\ref{eq:M}), taking into account contributions from both the short-range and long-range parts of the impurity potential. This derivation complements our previous studies of the phonon-induced spin relaxation and the inter-axis impurity-induced spin relaxation.\cite{Li_PRL11,Song_PRB12,Song_PRL14}   Importantly, the complete analytical framework in Sec.~\ref{sec:relax} allows one a fast calculation of the spin relaxation time in $n$-type silicon due to the interaction of electrons with phonons and impurities for any temperature, strain level, donor identity and concentration. Depending on the angle between strain and spin orientation, we have provided analytical expressions that quantify the anisotropy in the spin relaxation time. 

This work provides a clear motivation for employing silicon spintronic devices in which the spin transport region is compressively strained along one of the crystallographic axes. We have quantified the improvement in the spin relaxation of \textit{n}-type silicon when applying this type of strain. The spin relaxation time improves evidently when the strain is large enough to depopulate the electrons from the high-energy valleys. In non-degenerate silicon, this condition is met when $\Delta_v \gg k_BT$, where $\Delta_v$ is the strain-induced valley splitting energy. In degenerately doped silicon, the condition is met when $\Delta_v-\mu \gg k_BT$, where $\mu$ is the chemical potential. The results show that the inter-axis elastic impurity scattering is quenched more effectively by the strain compared with the inter-axis inelastic phonon scattering. As a result, a larger improvement is expected in degenerately doped Si:Sb compared with Si:P due to the relatively large spin-orbit coupling of Sb, and hence the bigger effect of the electron-impurity interaction on the spin relaxation in Si:Sb. We have predicted that the spin relaxation time  can be enhanced by nearly three orders of magnitude at low temperature Si:Sb, and by slightly more than one order of magnitude in Si:P under the same strain level, donor concentration, and temperature conditions. 

Another finding of our work is that the long-range interaction of electrons with the ionized impurity potential (Coulomb scattering) bears no practical significance for spin relaxation in silicon. It is much weaker than the inter-axis spin relaxation mechanisms and is also a weak effect when considering the contribution from all other intra-axis mechanisms (especially acoustic phonons or the short-range interaction with impurities in Si:Sb and Si:As). On the other hand, this intra-axis and long-range interaction is known to play a crucial role in limiting the electron mobility of doped semiconductors.\cite{Song_PRL14,Masetti_IEEE_83,Morin_PR54,Wolfstirn_JPCS60,Furukawa_LPSJ61,Granacher_JPCS67,Mousty_JAP74,Kaiblinger-Grujin_JAP98} Its insignificance for spin relaxation versus its importance for momentum relaxation implies that estimating the Elliott-Yafet spin relaxation time in \textit{n}-type silicon by multiplying the momentum relaxation time with some coefficient that depends on the spin-orbit coupling is an arbitrary choice.

\section*{ACKNOWLEDGMENTS} This work was supported by the Department of Energy under Contract No. DE-SC0014349, the National Science Foundation under Contract No. DMR-1503601, and the Defense Threat Reduction Agency under Contract No. HDTRA1-13-1-0013.

\appendix


\section{Structure of the $\mathbf{k}\cdot\mathbf{p}$ Hamiltonian} \label{sec:symmetries}
This Appendix includes details on the derivation of the $\mathbf{k}\cdot\mathbf{p}$ Hamiltonian, which we used throughout the main text.  Unlike many semiconductors in which the band extrema reside at highly symmetrical points of the Brillouin zone (e.g., the $\Gamma$-point in GaAs or the $L$ point in germanium), the conduction band minimum in silicon resides on the $\Delta$-symmetry axis, connecting the $\Gamma$ and $X$ points. The minimum is located about 0.15$(2\pi/a)$ away from the $X$ point ($k_0=0.15k_X$). In choosing between the symmetry groups of the $\Delta$-axis and the $X$-point to describe the electronic states at the bottom of the conduction bands in silicon, we choose the $X$-point since it has four-times more symmetry operations ($G^2_{32}$ versus $C_{4v}$). The set of  $X$-point symmetry operations is large-enough to determine a compact form of the $\mathbf{k}\cdot\mathbf{p}$ Hamiltonian, and we find that second-order perturbation theory in $k_0$ produces accurate eigensystem along the $\Delta$-axis, despite the fact that $k_0$ is not negligible.


\subsection{Time reversal symmetry}\label{sec:TRS}

We begin with invoking time-reversal argumentation in order to determine whether amplitudes in front of different terms in the Hamiltonian are real or imaginary. The time-reversal operator $\hat T=\sigma_y\hat K$, where $\hat K$ is the  operator of complex conjugation, anti-commutes with momentum:
\begin{equation}  \label{eq:3}
  \hat{\vec p}=-i\hbar \nabla  \Longrightarrow \{\mathbf{k}\cdot\mathbf{p},\hat T\}=0.
\end{equation}
From~\eqref{eq:8} we conclude that complex conjugation $\hat K$ in our basis is represented by $\rho_x$ for both $X_1$ and $X_4$. As shown in Tab.~\ref{tab:Hensel}, the same is true for the inversion $\hat I$, so for our basis functions these two operators are identical: $\hat K=\hat I$.
Time-reversal transforms momentum matrix elements as
\begin{equation}  \label{eq:10}  \begin{split}
    \langle \uparrow &X_i|\mathbf{k}\cdot\mathbf{p}|\uparrow X_j\rangle =(\hat T\mathbf{k}\cdot\mathbf{p}\uparrow X_j,\hat T\uparrow X_i)\\ =&-\rho_x(\downarrow X_i,\mathbf{k}\cdot\mathbf{p}\downarrow  X_j)^*\rho_x=\langle \downarrow X_i|\mathbf{k}\cdot\mathbf{p}|\downarrow X_j\rangle^*,
\end{split}\end{equation}
where the last equality is obtained by applying $\hat I$ on top of $\hat T$. Analogously
\begin{equation}  \label{eq:33}
    \langle \uparrow X_i|\mathbf{k}\cdot\mathbf{p}|\downarrow X_j\rangle =-\langle \downarrow X_i|\mathbf{k}\cdot\mathbf{p}|\uparrow X_j\rangle^*.
\end{equation}
In a similar fashion, one can transform matrix elements for operators that commute with $\hat T$. One of these operators is the spin-orbit interaction of the host crystal (which also commutes with $\hat I$),
\begin{equation}  \label{eq:11}  \begin{split}
\langle &\uparrow X_i|\orVomAt|\uparrow X_j\rangle =\\
&=-\langle \downarrow X_i|\orVomAt|\downarrow X_j\rangle^*,\\
\langle &\uparrow X_i|\orVomAt|\downarrow X_j\rangle =\\
&=\langle \downarrow X_i|\orVomAt|\uparrow X_j\rangle^*,
\end{split}\end{equation}
where $U_{\mathrm{at}}$ is the (intrinsic) atomic potential in pure silicon crystal.
Finally, we consider operators that represent the impurity potential $V$  and its spin-orbit interaction $\propto\orVomV$. They are not symmetric under inversion so their time-reversal symmetry relations contain $\rho_x$-matrices:
\begin{equation}  \label{eq:34}  \begin{split}
    \langle \uparrow X_i|\hat O|\uparrow X_j\rangle &=\rho_x\langle \downarrow X_i|\hat O|\downarrow X_j\rangle^*\rho_x,\\
    \langle \uparrow X_i|\hat O|\downarrow X_j\rangle &=-\rho_x\langle \downarrow X_i|\hat O|\uparrow X_j\rangle^*\rho_x,
\end{split}\end{equation}
where $\hat O=V$ or $\hat O=\orVomV$.

\begin{table}
\centering
\renewcommand{\arraystretch}{1.25}
\begin{tabular}{C| C C}
\qquad \qquad  & X_1\qquad \qquad& X_4   \\ \hline 
\{C_{2y} | 0\}   \qquad \qquad& -\rho_y \qquad \qquad& \rho_y  \\
\{S_{4z} | 0\}  \qquad \qquad& -\rho_y \qquad \qquad& -i\rho_z \\
\{ \hat{I} | \tau \}         \qquad \qquad& \rho_x   \qquad \qquad& \rho_x \\
\eta   \qquad \qquad & -\rho_0 \qquad \qquad& -\rho_0 \\
\end{tabular}
\caption{Unitary representation of the symmetry operations $C_{2y}$, $S_{4z}$, and $\hat{I}   $ of the $X$ point along the $z$-axis, $\mathbf{k}_{X} = (2\pi/a)(1,0,0)$. $\tau = (a/4)(1,1,1)$ denotes the non-primitive translation, and $\eta=\{ E | \mathbf{t} \} $ represents a primitive translation for which $\exp(i\mathbf{k}_{X}\mathbf{t}) =-1$. See Ref.~[\onlinecite{Hensel_PR65}] for more details.}
\label{tab:Hensel}
\end{table}


\subsection{Spatial symmetries}\label{sec:spaSym}
The structure of the $\mathbf{k}\cdot\mathbf{p}$ Hamiltonian is determined by matrix elements $\langle X_i|\hat O|X_j\rangle $ where $i,j=1,4$, and $\hat O$ can represent $\mathbf{k}\cdot\mathbf{p}$ terms,
impurity potential $\hat V\!,$ or spin-orbit interaction terms $\orVomAt$ and $\orVomV$.  Let $R_i$ and $R_j$ be IR matrices that represent some symmetry operation
$\hat g$ acting on $X_i$ and $X_j$, respectively. The transformation properties of matrix elements that involve $\hat O$ then follow
\begin{equation}
\langle \hat gX_i|\hat g\hat O{\hat g}^{-1}|\hat gX_j\rangle =R_i^*\langle X_i|\hat O|X_j\rangle R_j^T.\label{eq:6}
\end{equation}
Following Ref.~[\onlinecite{Hensel_PR65}] , the operations $C_2$, $S_4$, and $\hat{I}$  are sufficient in order to determine the form of the $\mathbf{k}\cdot\mathbf{p}$ Hamiltonian in the vicinity of the $X$-point. Table.~\ref{tab:Hensel} shows the generators of these operations for the chosen conduction ($X_1$) and valence ($X_4$) states in (\ref{eq:8}). 

Below, we demonstrate how the application of (\ref{eq:6}) and Tab.~\ref{tab:Hensel} are used to evaluate the non-vanishing interband spin-independent matrix elements of  $\hat O = \mathbf{k}\cdot\mathbf{p}$. Focusing first on the inversion operation $\hat I$ for which $R_1=R_4=\rho_x$, we obtain
\begin{equation}   \label{eq:23}
  \langle X_1|\,\mathbf{k}\cdot\mathbf{p}\,|X_4\rangle =-\rho_x\langle X_1|\,\mathbf{k}\cdot\mathbf{p}\,|X_4\rangle \rho_x,
\end{equation}
which means that $\langle X_1|\mathbf{k}\cdot\mathbf{p}|X_4\rangle $ cannot contain any terms proportional to $\rho_x$ or $\rho_0$. Similarly, we may use other symmetry operations to see further restrictions on $\langle X_1|\mathbf{k}\cdot\mathbf{p}|X_4\rangle $. We see that the terms $k_z\cdot (\rho_y,\rho_z)$ are forbidden by $S_4^2$, and $(k_x\cdot \rho_z,k_y\cdot \rho_y)$ by $C_2$. Finally, we are left with the spin-conserving invariant, $H_{cv} \propto ak_x\rho_y+ik_y\rho_z$. By applying $S_4$ on this expression, we realize that $a=1$.  The spin-independent part of $\langle X_1|\mathbf{k}\cdot\mathbf{p}|X_4\rangle $ is then given by
\begin{equation}  \label{eq:32}
   \langle X_1| \, \mathbf{k} \cdot \mathbf{p}\, |X_4\rangle =-iP(k_x\rho_y+ik_y\rho_z).
\end{equation}
Thus, using spatial symmetries we are able to determine that ~\eqref{eq:32} represents the spin-independent structure of the off-diagonal (interband) block in our impurity-free Hamiltonian. In order to see whether the coefficient $P$ in~\eqref{eq:32} is real or imaginary, we will have to employ time reversal symmetry. For the case of~\eqref{eq:32}, Eq.~\eqref{eq:10} for $i=1$ and $j=4$ makes it clear that the coefficient $P$ in~\eqref{eq:32} is real.

We can apply a similar approach to check which of the spin-dependent terms are symmetry forbidden. In the absence of impurities, such terms could arise from matrix elements of $\orVomAt$. We now have to consider not only the orbital parts of $R_i$ and $R_j$ in~\eqref{eq:6}, but their spinor-rotation parts as well. In spinor space, the generators of $\hat I$, $C_2$, and $S_4$ are represented by
\begin{equation}\label{eq:15}  \begin{split}
    S(\hat I)=-\sigma_0,\quad S(C_2)=i\sigma_y,\\ S(S_4)=\frac{\sigma_0-i\sigma_z}{\sqrt2}=\sqrt{-i\sigma_z}.
\end{split}\end{equation}
where $\sigma_0$ is a $2\times 2$ unity matrix acting in spinor space (analogous to $\rho_0\equiv\mathds1$, which denotes a $2\times 2$ ``orbital'' unity matrix). Considering first terms that are proportional to $\rho_0$, we see that $\hat I$ allows $(\sigma_0,\sigma_x,\sigma_y,\sigma_z)$, $C_2$ allows $(\sigma_0,\sigma_y)$, and $S_4^2$ allows $(\sigma_0,\sigma_z)$. Thus,  $\rho_0\sigma_0$  is the only allowed invariant term. Similarly, we deduce that $\rho_y\otimes\sigma_0$ is forbidden by $\hat I$.


\subsection{Intrinsic terms: $\langle X_i|\mathbf{k}\cdot\mathbf{p}|X_j\rangle $  and $\langle X_i|\orVomAt|X_j\rangle $}\label{sec:intrinsic}
The previous section showed how the method of invariants can be used to determine the general form of a $\mathbf{k}\cdot\mathbf{p}$ Hamiltonian.\cite{Bir_Pikus_Book,Winkler_Book} Following this method, we have previously derived the spin-dependent Hamiltonian near the $X$ point of diamond crystal structure, \cite{Song_PRB12}  finding that
\begin{eqnarray}  \label{eq:46}
  H_X &=& \begin{pmatrix} H_c&H_{cv}\cr H_{cv}^\dag &H_v \end{pmatrix}, \nonumber \\
H_c &=&\frac{\hbar^2k^2_{\perp}}{2m_{\perp}}+\frac{\hbar^2k^{'2}_z}{2m_z}+\hbar v_ck'_z\rho_z,  \\
H_v &=&\frac{\hbar^2k^2_{\perp}}{2m_{\perp}}+\frac{\hbar^2k^{'2}_z}{2m_z}, \nonumber \\
H_{cv}&=& -iP(k_x\rho_y+ik_y\rho_z)+i\Delta_X(\rho_x\otimes\sigma_y-\rho_0\otimes\sigma_x) \nonumber \\ 
&+& \alpha k_z'(i\rho_z\otimes\sigma_x+\rho_y\otimes\sigma_y)+ \alpha \sigma_z \nonumber
\end{eqnarray}
Here, $\mathbf{k}' \equiv ({\mathbf{k}}_{\perp},k_z')$, where $k_z'=k_X-k_z$ reaches zero at the $X$-point. The wavevector component away from the $\Delta$-axis, $\mathbf{k}_{\perp}$, is assumed small enough to be treated perturbatively.  While the intraband diagonal blocks, $H_c=\langle X_1|\mathbf{k}\cdot\mathbf{p}|X_1\rangle $ and $H_v=\langle X_4|\mathbf{k}\cdot\mathbf{p}|X_4\rangle $, are spin-independent  (diagonal in spinor space), the interband off-diagonal blocks are affected both by $\mathbf{k}\cdot\mathbf{p}$ and spin-orbit coupling terms. The latter gives rise to two independent crystal parameters,  $\Delta_X$ and $\alpha$, as we explain after (\ref{eq:21}) and (\ref{eq:a1}) in the main text. Time-reversal analysis from App.~\ref{sec:TRS} demonstrates that the coefficients $v_c$, $P$, $\Delta_X$, and $\alpha $ are real. We emphasize that the resulting form of the Hamiltonian is determined by the representation used in Tab.~\ref{tab:Hensel}. A different basis choice would result in a different set of generators, and consequently, in a different Hamiltonian form.


\subsection{Extrinsic terms: $\langle X_i|\hat V|X_j\rangle $ and $\langle X_i|\orVomV|X_j\rangle $}\label{sec:extrinsic}

The presence of impurities results in additional perturbative terms. The complete Hamiltonian for a silicon crystal with randomly placed impurities can be written in the mixed momentum-coordinate representation  as
\begin{equation}\label{disModel}
H(\vec\rho, \mathbf{k})=H_X(\mathbf{k})+\delta H(\mathbf{k})\!\sum_{j=1}^N\chi (\vec\rho-\!{\vec\rho}_{j})
\end{equation}
where  $N=N_D|\Omega |$  is the total number of (randomly placed) defects. $H_X(\mathbf{k})$ is the impurity-free $\mathbf{k}\cdot\mathbf{p}$ Hamiltonian, provided in \eqref{eq:46}, and $\vec\rho_j$ denotes the discrete coordinate that lists all of the cells with impurities. The function $\chi$ has a value of one within unit cells that contain impurities and zero otherwise, 
\begin{equation}\label{modelBesp}
\chi (\vec\rho\,)\!=\! \begin{cases}1, &\vec\rho\in v,\\0, &\vec\rho\notin v.\end{cases}
\end{equation}

In order to evaluate the form of $\delta H(\mathbf{k})$ in \eqref{disModel}, we note that the impurity potential, $V(\mathbf{r})$, has two contributions. The first one is invariant with respect to all operations of the diamond point group, while the second contribution transforms according to the IR $M_2'$ of the $G^2_{32}$ group. The latter flips sign under space inversion operation (transforms as  the product $xyz$). Repeating the invariant-based analysis in Apps.~\ref{sec:TRS} and \ref{sec:spaSym}, we obtain the following impurity-induced  interband correction terms (responsible for the Yafet process),
\begin{eqnarray} \label{eq:interband_imputity_correction}
    \delta H_{cv}&=&\frac1{|v|}\int_vX_1^*(\vec r\,)[V(\vec r\,)+\mathbf{k}\cdot\mathbf{p}]X_4(\vec r\,) \text{d}^3r\nonumber\\
&=&-iV_{cv}(\rho_y\otimes\sigma_x+i\rho_z\otimes\sigma_y)  \label{Ui}\\
&+&V_{i1}k_z\rho_x\otimes\sigma_0+V_{i2}(\rho_xk_y-k_x\rho_0)\otimes\sigma_0,\nonumber
\end{eqnarray}
where the X-basis functions are normalized with respect to the elementary cell volume~$|v|$,
\begin{equation}\label{eq:36}  \begin{split}
    ||X_1||_v^2=||X_4||_v^2\equiv \int_v|X_4(\vec r\,)|^2\text{d}^3r=|v|. 
\end{split}\end{equation}

A different elegant way to derive the interband invariants in (\ref{eq:interband_imputity_correction}) is to make use of the selection rules of $M_2'$ with transverse vector components ($k_x$ and $k_y$) and pseudovector ones ($\sigma_x$ and $\sigma_y$). Transformation properties of the former/latter are represented by $M_5/M_5'$, and their interaction with $M_2'$ flips their roles,
\begin{eqnarray} \label{tum}
M_2' \times M_5 = M_5' \qquad  \qquad  M_2' \times M_5' = M_5 \,\,.
\end{eqnarray}
That is, $M_2'$ switches between the transverse components of vectors ($M_5$) and pseudovectors ($M_5'$):  $x \,\leftrightarrow\, \sigma_x$ and $y \,\leftrightarrow\, \sigma_y$. This exchange rule establishes a connection between linear in momentum terms from~\eqref{eq:46} with impurity-induced spin-flip corrections to these equations. For example, let us first inspect the invariant  $k_x\rho_y+ik_y\rho_z$ in~\eqref{eq:46}, which contains the $x$ and $y$ vector components of $\mathbf{k}$ ($M_5$ IR). In the disorder-induced part of the Hamiltonian, $\delta H$, impurity  terms transform according to $M_2'$. Therefore, we should replace the vector components in $k_x\rho_y+ik_y\rho_z$ with pseudovector ones in order to find the analogous term in $\delta H$. The only pseudovector we have is spin, so that $\delta H$ includes the invariant $\rho_y\otimes\sigma_x+i\rho_z\otimes\sigma_y$ (corresponding to the first term in (\ref{eq:interband_imputity_correction})). 

In order to compare the dominant interband spin-mixing terms in \eqref{eq:46} and \eqref{eq:interband_imputity_correction}, we make use of the fact that $ V_{cv}\sim \delta_B \Delta_{so}$, where $\delta_B = a_B^3/V$ and $\Delta_{so}$ is the spin-splitting of the impurity ground state. We get that 
\begin{equation}  \label{eq:43}
 \renewcommand{\arraystretch}{1.5}
\begin{tabular}{c c c c} 
& \qquad Si:P\qquad &\qquad Si:As\qquad &\qquad Si:Sb  \\ \hline   
 $V_{cv}$(meV)$\approx $ & \qquad 3.0 \qquad& \qquad 10.6 \qquad& \qquad 31.3 
\end{tabular}\end{equation}
We can now compare the interband spin-mixing amplitude of the impurity with that of the host atoms  ($V_{cv}$ in~\eqref{eq:interband_imputity_correction}  versus $\Delta_X$ in~\eqref{eq:46}). The latter gives rise to the interband spin-mixing in clean silicon at $k'=0$. Given that Si and P are neighboring elements in the periodic table, it is reasonable that their spin-orbit coupling parameters are comparable in Si:P,  $V_{cv}=3.0$\,meV and $\Delta_X=3.6$\,meV.

Turning to impurity-induced intraband terms, we follow the analysis of App.~\ref{sec:spaSym} in order to find the spin-dependent correction of $H_c$. This correction term is responsible for the central-cell Elliott process,
\begin{equation} \label{dHU}
\delta H_c=\delta_B\Delta_0\rho_0\otimes\sigma_0+\delta_B\Delta_1\rho_y\otimes\sigma_0 \,,
\end{equation}
where $\delta_B\Delta_0$ denotes short-range spherically-symmetric corrections to the screened Coulomb potential.\cite{Ralph_PRB75} The second term in~\eqref{dHU}, $\delta_B\Delta_1\rho_y$, arises due to the low-symmetry ($T_d$) part of the impurity potential, which is responsible for the spin-independent splitting of the $s$-state into singlet, triplet, and doublet. Finally, the  intra- valence band impurity terms follow
\begin{equation}   \label{eq:9}
\delta H_v=V_v\rho_z\otimes\sigma_z,
\end{equation}
which partially lifts the degeneracy of the valence band near the $X$ point.


\begin{widetext}
\section{Correction factors due to the intervalley electron-phonon interaction} \label{App:Fterms} 
The general expressions for the three intervalley $f$-process correction factors,  $C_{\Sigma_{1-3}}$, that appear in (\ref{tau_inter_axis_strain}) follow
\begin{eqnarray}  \label{eq:Cf_phonon_general}
C_{\Sigma_j}  &=& \frac{3}{\mathcal{I}_1(0)+2\mathcal{I}_1( \Delta_v)} \cdot \frac{1}{A_j r_j K_1(r_j)}  \times  \Big\{  \left[   2e^{-r_j}\mathcal{I}_{+,j}(\Delta_v,\Delta_v-\varepsilon_j) + 2e^{r_j}\mathcal{I}_{-,j}(\Delta_v+\varepsilon_j,\Delta_v)   \right] (1 + B_j\delta_{s,v} )  \nonumber  \\ & + & \left[   e^{-r_j}\left( \mathcal{I}_{+,j}(\Delta_v,-\varepsilon_j)+ \mathcal{I}_{+,j}(0,\Delta_v-\varepsilon_j)\right) +  e^{r_j}\left( \mathcal{I}_{-,j}(\Delta_v,\varepsilon_j)+ \mathcal{I}_{-,j}(\Delta_v+\varepsilon_j,0)\right) \right](D_j - R_j\delta_{s,v} )   \Big\}, 
\end{eqnarray}
where $\{A_j, B_j, D_j, R_j \}$ are respectively $\{4, -1, 1, -1\}$ for the $\Sigma_1$ mode, and $\{8, 1, 3, 1\}$  for the $\Sigma_2$ and $\Sigma_3$ modes.  In addition, $r_j = T_j/2T$ and $\varepsilon_j = k_BT_j$, while $T_1=540$~K, $T_2=660$~K, and $T_3=270$~K. Terms that include $e^{r_j}$ ($e^{-r_j}$) are associated with phonon emission (absorption). The integral form of $\mathcal{I}_1$ is provided in (\ref{tau_inter_imp_IN}).  The other integral forms are defined as
\begin{eqnarray}  \label{tau_If_pm_j}
\mathcal{I}_{\pm,j}(\varepsilon_a,\varepsilon_b) = \int_{\varepsilon_m}^\infty \!  d\varepsilon \frac{\sqrt{(\varepsilon-\varepsilon_a)(\varepsilon-\varepsilon_b)}}{k_BT}     \left[  \,\,  \frac{\partial \mathcal{F}}{\partial \varepsilon}\Big|_{\varepsilon} +  \mathcal{F}(\varepsilon) \frac{\partial \mathcal{F}}{\partial \varepsilon}\Big|_{\pm \varepsilon_j}  - \mathcal{F}(\pm \varepsilon_j) \frac{\partial \mathcal{F}}{\partial \varepsilon}\Big|_{\varepsilon}  \,\, \right] \,,
\end{eqnarray}
where $\varepsilon_m = \text{max}\{\varepsilon_a,\varepsilon_b\}$. The second and third terms in square parentheses stem from the inelastic nature of the scattering ($\varepsilon_j \neq 0$) and they play a role when deviating from the Boltzmann regime. Figures~\ref{fig:cf}(b)-(d) show the numerical values of  $C_{\Sigma_{1-3}}$ as a function of valley splitting energy ($\Delta_v$) for various temperatures and donor concentrations. In the Boltzmann limit, we get 
\begin{eqnarray}  \label{inter_axis_F_phonon_Boltzmann}
C_{\Sigma_j}  &\rightarrow& \frac{3e^{-r_v}}{1+2e^{-2r_v}} \Bigg\{ \frac{2e^{-r_v}}{A_j} (1 + B_j\delta_{s,v}) + \frac{|r_v \!-\! r_j| K_1(|r_v \!-\! r_j|) + (r_v \!+\! r_j)K_1(r_v \!+\! r_j) }{A_jr_j K_1( r_j )}(D_j - R_j\delta_{s,v} ) \Bigg\},\,\,\,\, 
\end{eqnarray}
where  $r_v = \Delta_v/2k_BT$ and $K_1(x)$ is the first-order modified Bessel function of the second kind.

The general expression for the intervalley $g$-process correction factor that appears in (\ref{tau_intra_axis_strain}) follows
\begin{eqnarray}  
C_{\Delta}  = \frac{3}{\mathcal{I}_1(0)+2\mathcal{I}_1( \Delta_v)} \cdot \frac{1}{8r_g^2 K_2(r_g)}  \times  \Big\{  \!\!\!   &&  \!\! \left[   e^{-r_g}\mathcal{I}_{+,g}(\Delta_v,\Delta_v-\varepsilon_{\Delta_1}) + e^{r_g}\mathcal{I}_{-,g}(\Delta_v,\Delta_v+\varepsilon_{\Delta_1})   \right] (3 - \delta_{s,v} )   \nonumber  \\  && \!\! +  \left[   e^{-r_g} \mathcal{I}_{+,g}(0,-\varepsilon_{\Delta_1})+ e^{r_g}\mathcal{I}_{-,g}(0,\varepsilon_{\Delta_1}) \right]( 1 + \delta_{s,v} ) \,\,  \Big\}, 
\end{eqnarray}
where $r_g = T_g/2T$ and $\varepsilon_{\Delta_1} = k_BT_g$ ($T_g=240$~K). In addition,
\begin{eqnarray}  \label{tau_Ig_pm_j}
\mathcal{I}_{\pm,g}(\varepsilon_a,\varepsilon_b) = \int_{\varepsilon_m}^\infty \!  d\varepsilon \left[ \frac{\sqrt{\varepsilon-\varepsilon_a}}{k_BT} \left( \frac{\varepsilon-\varepsilon_b}{k_BT} \right)^{\!\frac{3}{2}} +    \frac{\sqrt{\varepsilon-\varepsilon_b}}{k_BT} \left( \frac{\varepsilon-\varepsilon_a}{k_BT} \right)^{\!\frac{3}{2}}     \right] \left[  \,\,  \frac{\partial \mathcal{F}}{\partial \varepsilon}\Big|_{\varepsilon} +  \mathcal{F}(\varepsilon) \frac{\partial \mathcal{F}}{\partial \varepsilon}\Big|_{\pm \varepsilon_{\Delta_1}} \!\!\!\! - \mathcal{F}(\pm \varepsilon_{\Delta_1}) \frac{\partial \mathcal{F}}{\partial \varepsilon}\Big|_{\varepsilon}  \,\, \right],
\end{eqnarray}
where $\varepsilon_m = \text{max}\{\varepsilon_a,\varepsilon_b\}$
In the Boltzmann limit the expression for $C_{\Delta}$ becomes compact and provided by (\ref{tau_C_intra_axis_phonon_Bolt}).

\end{widetext}

\end{document}